# Crystal-surface polarity enables nonreciprocal transport in a centrosymmetric kagome magnet

Daiki Yamaguchi[1,2*], Max T. Birch[2], Hibiki Onodera[1], Akiko Kikkawa[2], Naoto Nagaosa[2,3], Yoshinori Tokura[1,2,4*]

[1]*Department of Applied Physics, The University of Tokyo, Tokyo 113-8656, Japan*

[2]*RIKEN Center for Emergent Matter Science (CEMS), Wako 351-0198, Japan*

[3]*Fundamental Quantum Science Program (FQSP), TRIP Headquarters, RIKEN, Wako 351-0198, Japan*

[4]*Tokyo College, The University of Tokyo, Tokyo 113-8656, Japan*

DY*: dai20yama011-1@g.ecc.u-tokyo.ac.jp,

MB: maximilian.birch@riken.jp,

HO: onodera-hibiki714@g.ecc.u-tokyo.ac.jp,

AK: kikkawa@riken.jp,

NN: nagaosa@riken.jp,

YT*: tokura@riken.jp

*Correspondence should be addressed to D.Y. or Y.T.

**Abstract**

Every crystal surface breaks inversion symmetry even when the bulk is centrosymmetric, creating an intrinsic boundary polarity (***P***). Here we show that this polarity enables zero-field nonreciprocal transport in the centrosymmetric kagome magnets $TbMn_6Sn_{6-x}In_x$. In symmetric $TbMn_6Sn_{4.4}In_{1.6}$ microdevices, opposing side surfaces exhibit second-harmonic longitudinal resistances of comparable magnitude and opposite sign. Reversing the spontaneous magnetization (***M***) reverses both signals, whereas rotating ***M*** so that the toroidal moment ***P*** × ***M*** has no component along the current suppresses the response. In pristine $TbMn_6Sn_6$, the signal switches off across the spin-reorientation transition, providing a chemically ordered control of the same symmetry relation. Spatial potential mapping indicates that the nonlinear boundary contribution is distinct from the bulk linear conduction. Spin-orbit-coupled surface states associated with topological bands may enhance the effect. Crystal-surface polarity therefore offers a broadly applicable route to zero-field rectification in centrosymmetric magnetic conductors around room temperature.

**Introduction**

Nonreciprocal electronic transport distinguishes opposite current directions and requires inversion-symmetry breaking[1,2]. In a magnetic conductor, simultaneous breaking of inversion and time-reversal symmetries permits a second-order longitudinal response. At the phenomenological level, the relevant symmetry can be represented by a toroidal moment ***T*** proportional to ***P*** × ***M*** , where ***P*** is a polar vector and ***M*** is the magnetization[3,4]. Coupling between ***T*** and the current density ***j*** then allows a nonreciprocal resistance when either ***P*** or ***M*** is reversed[1,2,5-8]. Although symmetry determines whether this response is allowed, its magnitude is set by the electronic structure and scattering processes.

Materials with strong spin-orbit coupling are especially promising because their spin texture and quantum geometry can amplify nonlinear transport.

Most demonstrations have relied on intrinsically polar or chiral crystals, or on deliberately patterned noncentrosymmetric structures. A recent example used nanosculpted three-dimensional helices of the centrosymmetric magnetic Weyl semimetal $Co_3Sn_2S_2$[8]. The helical geometry imposed a polarity perpendicular to the spontaneous magnetization and produced zero-field nonreciprocity, with the sign controlled by the handedness. This approach established geometrical symmetry engineering, but it required a specialized three-dimensional device and unequal inner and outer surfaces to avoid cancellation.

A more general source of polarity is present at every crystal boundary: a surface necessarily distinguishes the outward and inward directions even when the bulk is centrosymmetric. Boundary symmetry reduction also underlies chiral edge states in quantum Hall systems, topological-insulator surface states and Fermi arcs in Weyl semimetals[9-13]. Nonlinear transport has begun to reveal boundary contributions in magnetic topological-insulator heterostructures[14] and surface-specific Fermi-arc transport in noncentrosymmetric TaAs under an applied field[15]. It remains unknown whether the intrinsic polarity of an ordinary crystal surface can itself generate a zero-field nonreciprocal response in a centrosymmetric magnetic metal, and whether that response can be switched by a bulk magnetic phase transition.

Here we address these questions in the centrosymmetric kagome ferrimagnets $TbMn_6Sn_6$ and $TbMn_6Sn_{4.4}In_{1.6}$. A symmetric Hall-bar device resolves nonlinear voltages independently on two opposing side surfaces, whose polarities are equal in magnitude and opposite in direction. In $TbMn_6Sn_{4.4}In_{1.6}$, the nonreciprocal signals on side surfaces have comparable magnitude and opposite sign, reverse with the spontaneous magnetization and remain finite without an applied field. In pristine $TbMn_6Sn_6$, the response is switched off when the magnetization rotates at the spin-reorientation transition. These results establish crystal-surface polarity, combined with magnetic order, as a general materials principle for nonreciprocal boundary transport. Topological surface bands may enhance the magnitude but are not assumed as the symmetry-breaking origin.

## Results

### Magnetic platform for boundary polarity

$TbMn_6Sn_6$ crystallizes in the centrosymmetric $HfFe_6Ge_6$-type hexagonal structure (space group P6/mmm, no. 191), with Mn kagome layers stacked along the *c*-axis (Fig. 1a)[16]. Antiparallel Tb and Mn moments form a ferrimagnetic state[17]. Upon cooling, the parent compound first orders into an easy-plane state below the Curie temperature $T_C \approx 420$ K and then rotates into an easy-axis state below the spin-reorientation transition temperature $T_{SR} \approx 315$ K (Fig. 1b,c)[18,19]. Its kagome-derived bands, out-of-plane magnetization and strong spin-orbit coupling produce gapped Dirac bands and multiple three-dimensional topological crossings near the Fermi energy[20-23]. Their surface projections can support Fermi-arc-like states with opposite propagation characteristics on opposing side surfaces (Fig. 1f,g). We use this electronic structure as a possible microscopic amplifier of the boundary response, while the symmetry argument below does not rely on assigning a particular surface band.

Single crystals of $TbMn_6Sn_{4.4}In_{1.6}$ were grown by the laser floating-zone method (Methods and Supplementary Fig. S1). Substitution expands the *c*-axis while changing the *a*-axis only weakly. More importantly, it lowers $T_C$ to 296 K and removes the spin-reorientation transition, leaving out-of-plane ferrimagnetism throughout the ordered phase[24,25] (Fig. 1d). This magnetic state is advantageous because, for current along $x \parallel \langle 100 \rangle$ and a side-surface polarity along $y \parallel \langle 120 \rangle$, the longitudinal toroidal coupling is proportional to $(\boldsymbol{P} \times \boldsymbol{M}) \cdot \boldsymbol{j} = P_y M_z j_x$. Single crystals of pristine $TbMn_6Sn_6$ were grown by Sn self-flux method (Methods and Supplementary Fig. S1). Microdevices of $TbMn_6Sn_{4.4}In_{1.6}$ and $TbMn_6Sn_6$ were fabricated by the focused-ion-beam (FIB) milling technique (Methods and Supplementary Fig. S2). The microdevice resistivities have similar temperature dependences at high temperature, although In substitution raises the residual resistivity to about 100 μΩcm (Fig. 1e). The bulk and microdevice resistivities have similar temperature dependences and transition temperatures (Supplementary Fig. S2), indicating no severe damage during the FIB milling. Magnetoresistance and anomalous Hall measurements further confirm that FIB fabrication preserves the characteristic magnetic transport properties (Supplementary Fig. S3).

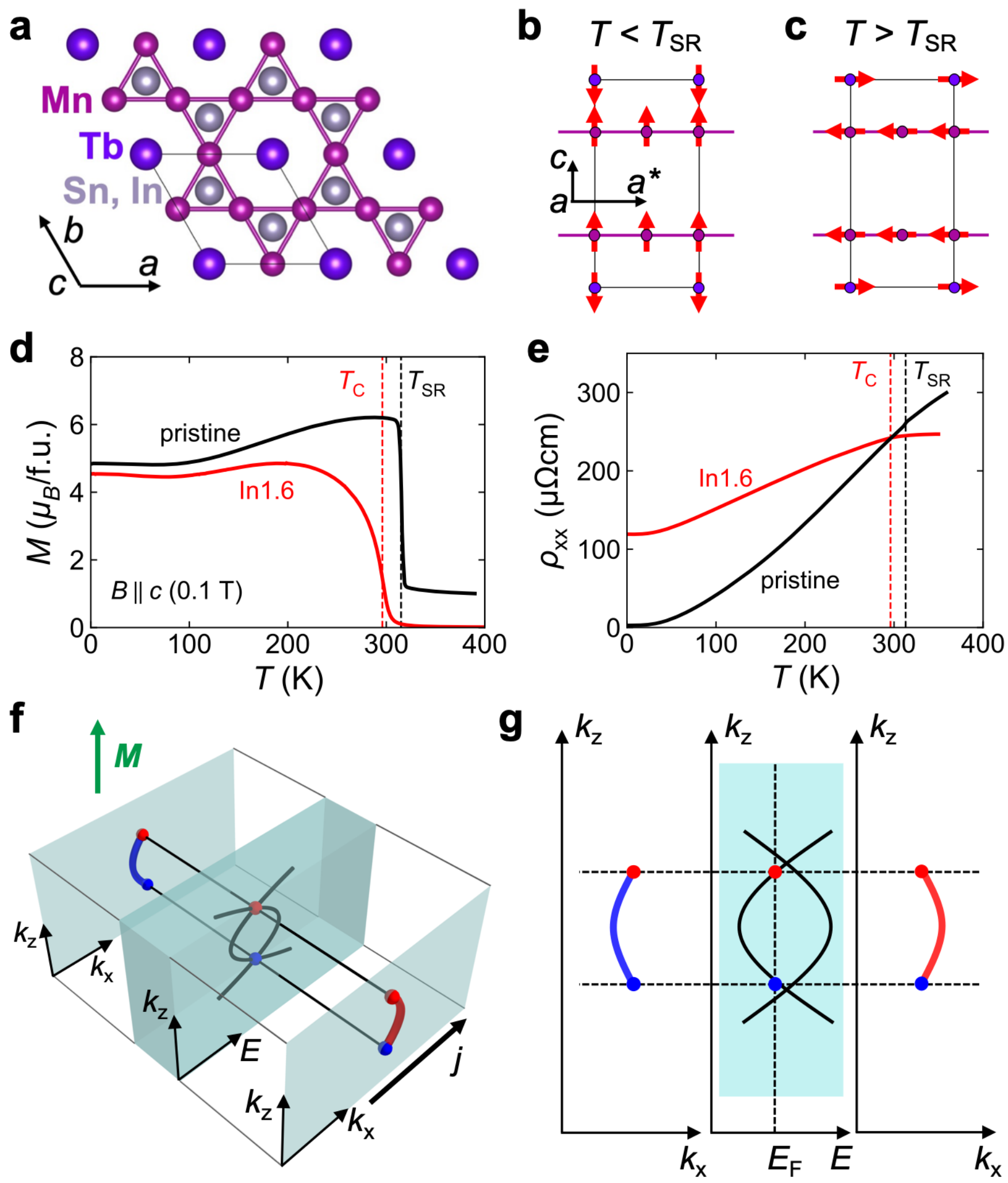

**Fig. 1: Crystal structure and characterization of $TbMn_6Sn_{4.4}In_{1.6}$ and $TbMn_6Sn_6$.**

**a**, Crystal structure viewed along the crystallographic *c*-axis, drawn using VESTA[28]. Purple, magenta and grey spheres denote Tb, Mn and Sn/In sites, respectively; grey lines outline the unit cell. **b, c,** Out-of-plane (**b**) and in-plane (**c**) ferrimagnetic structures viewed along the *a*-axis. Red arrows denote the Tb and Mn moments. Sn/In sites are omitted for clarity. **d**, Temperature dependence of the magnetization of In-substituted $TbMn_6Sn_{4.4}In_{1.6}$ (red) and pristine $TbMn_6Sn_6$ (black), measured with $B \parallel c$ at 0.1 T after field cooling from 390 K to 2 K. **e,** Temperature dependence of the in-plane longitudinal resistivity $\rho_{xx}$ for focused-ion-beam (FIB) fabricated microdevices of $TbMn_6Sn_{4.4}In_{1.6}$ (red) and $TbMn_6Sn_6$ (black). Red and black vertical dashed lines in (**d**) and (**e**) denote $T_C$ of the In-substituted compound and $T_{SR}$ (spin-reorientation transition temperature) of the pristine compound, respectively. **f**, Schematic surface states generated by bulk band crossings. A representative bulk dispersion along $k_z$ is shown in the centre. Red and blue points mark Weyl nodes of opposite chirality, and the thick red and blue curves in the opposing $k_x$-$k_z$ surface planes represent Fermi arcs of opposite orientation. In the nonreciprocal transport measurement configuration, the magnetization $\boldsymbol{M}$ is along $z$ and the ac current $\boldsymbol{j}$ flows along $x$. **g,** Schematic Fermi arcs at the two opposing side surfaces when $E_F$ does not coincide with the Weyl nodes. Red and blue dots denote the points at which the surface arc merges with the bulk bands.

**Opposite responses at opposing side surfaces**

We fabricated eight-terminal Hall bars from $TbMn_6Sn_{4.4}In_{1.6}$ single crystals by FIB milling (Fig. 2a). The $c$-axis is normal to the device plane, so the out-of-plane magnetization $\boldsymbol{M}$ is perpendicular to the polarity $\boldsymbol{P}$ of each side surface. The two surfaces therefore carry toroidal moments $\boldsymbol{T}$ proportional to $\boldsymbol{P} \times \boldsymbol{M}$ of equal magnitude and opposite direction (Fig. 2b,c). At 260 K, the field-antisymmetric longitudinal second-harmonic (2$f$) voltages on the left and right sides are hysteretic, remain finite at zero field and have nearly equal magnitudes with opposite signs (Fig. 2f,g). The magnitude of these signals is comparable possibly because of the precise FIB device fabrication of the voltage contacts. Reversing $\boldsymbol{M}$ reverses both signals. Rotating $\boldsymbol{M}$ so that the longitudinal component of $\boldsymbol{P} \times \boldsymbol{M}$ vanishes suppresses the response (Supplementary Fig. S8). This means, when the magnetic field is parallel or anti-parallel to the side surface polarities, contributions from both the boundary and the bulk polarities are not discernible. From this experimental fact, the bulk centrosymmetry is shown to be maintained even in the In-substituted crystal. A three-pair transverse-voltage map provides a spatial check: the two outer pairs reproduce the opposite side signals, whereas the central pair is nearly zero (Fig. 2d,e,h,i). This is a robust demonstration that the apparent nonlinear Hall signal is a consequence of the opposite nonreciprocal longitudinal transport on the two side surfaces. By contrast, the first-harmonic (1*f*) longitudinal voltage is the same on both sides (see Fig. 3d). The combined sign, magnetization and terminal dependences identify a boundary-odd nonlinear contribution superposed on bulk-dominated linear conduction.

Extended demonstration of 2*f* potential mapping along a longer microdevice with 7 voltage terminals on one side is provided in Supplementary Fig. S9. The result is consistent with the result in Fig. 2. Since $TbMn_6Sn_6$ exhibits a large thermoelectric response[21,22], extrinsic Joule-heating effect (potentially the Nernst-effect induced 2*f* voltage signal) might be able to explain the present results. However, it is ruled out in the discussion in Supplementary Note 1, utilizing the result of the extended potential mapping and also the frequency independence of the signal up to 10 kHz (around this frequency, there seems to be an upper limit for the proper ac excitation input to a microdevice) shown in the Supplementary Fig. S10.

The sign reversal between the two surfaces follows from symmetry and does not by itself identify a unique microscopic channel. A minimal chiral-surface-band model nevertheless illustrates how such a response can arise. For each $k_z$, we expand the dispersion near a surface Fermi point as

$$\xi_k = v_F k + \alpha k^2, \qquad (1)$$

where $k$ is measured from the Fermi point and the nonlinear terms encode particle-hole asymmetry. Expanding the steady-state Boltzmann distribution to second order in the electric field $E$ gives

$$f_2(k) = \left(\frac{eE\tau}{\hbar}\right)^2 \frac{d^2 f_0(k)}{dk^2}, \qquad (2)$$

where $f_0(k)$ is the equilibrium distribution and $\tau$ is the relaxation time. The second-order conductivity is then

$$\sigma_2 = \frac{e^3 \tau^2 \alpha}{\pi \hbar^3} \mathrm{sign}(v_F). \qquad (3)$$

Inversion maps one side surface to the other and reverses the chiral dispersion, so the second-order conductivity $\sigma_2$ changes sign. The model is compatible with a contribution from Fermi-arc-like states, but the transport data do not uniquely distinguish those states from other spin-orbit-coupled surface bands. The toroidal surface symmetry dictates the existence and sign of the edge nonreciprocity, whereas the quantum geometric Berry curvature of the spin-orbit-coupled topological surface states may provide a microscopic route to its pronounced magnitude.

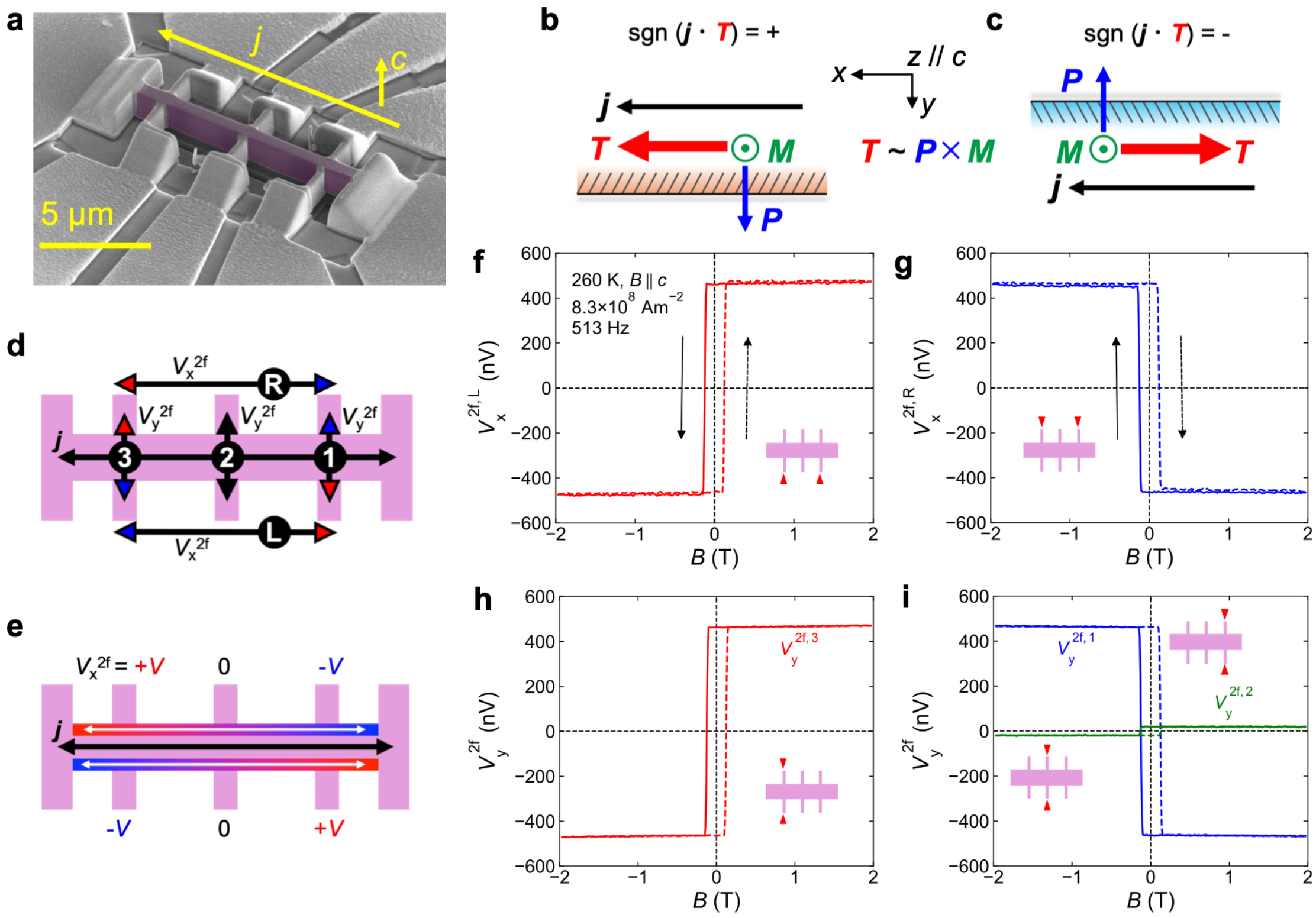

**Fig. 2: Side-surface asymmetric nonreciprocal transport in a $TbMn_6Sn_{4.4}In_{1.6}$ microdevice at 260 K.**

**a**, Scanning electron microscopy image of the FIB fabricated device; the sample body is shaded purple. **b**, **c**, Definition of the toroidal moment $\boldsymbol{T} \propto \boldsymbol{P} \times \boldsymbol{M}$ at the left (**b**) and right (**c**) side-surfaces. $\boldsymbol{P}$ and $\boldsymbol{M}$ denote the surface polarity and out-of-plane spontaneous magnetization, respectively. The schematics are viewed along the *c*-axis; hatched solid lines indicate the side walls. The left and right sides relative to the current direction are shaded orange and light blue, respectively. Black arrows indicate the ac current density $\boldsymbol{j}$. **d**, Measurement configuration. The second-harmonic (2*f*) longitudinal voltages are measured simultaneously on the left (L) and right (R) sides, and the 2*f* transverse voltages are measured using terminal pairs 1-3. **e**, Schematic 2*f* potential distribution, with opposite signs on the two sides. **f**, **g**, Magnetic-field dependence of the 2*f* longitudinal voltages $V_x^{2f,\,L}$ (**f**) and $V_x^{2f,\,R}$ (**g**). **h**, **i**, Magnetic-field dependence of the 2*f* transverse voltages $V_y^{2f,\,3}$ (**h**) and $V_y^{2f,\,1}$ and $V_y^{2f,\,2}$ (**i**). The field is applied along the *c*-axis, and the ac electric current of $j = 8.3 \times 10^8$ Am$^{-2}$ at $f = 513$ Hz is applied. The solid and dashed curves correspond to data taken in decreasing and increasing field, respectively, as indicated by the black arrows in (**f**) and (**g**).

**Temperature-dependent nonreciprocity**

We define the nonreciprocal resistance as $R_{xx}^{2f} = V_x^{2f}/I$, where $I$ is the applied current. As expected for a second-order voltage, $R_{xx}^{2f}$ is proportional to $I$ and hence to the current density $j$ (Supplementary Fig. S5). The left and right signals retain predominantly opposite signs throughout the ordered phase (Fig. 3a-c, Supplementary Fig. S6). At zero field, their hysteretic component grows upon warming and disappears at $T_C$ when spontaneous ferrimagnetism is lost. Under a magnetic field, the finite response persists even above $T_C$ because the field induces an out-of-plane magnetization. The temperature dependence of the nonlinear signal differs markedly from those of both the anomalous Hall resistivity and the first-harmonic longitudinal resistivity (Fig. 3d), showing that it is not a simple replica of bulk linear transport (see also Supplementary Note 2). A smaller side-symmetric component and a low-temperature sign change vary between surfaces (dashed rectangular region in Fig. 3c), plausibly reflecting surface-dependent elastic scattering; the robust result is the boundary-antisymmetric component $[R_{xx}^{2f,\,L} - R_{xx}^{2f,\,R}]/2$.

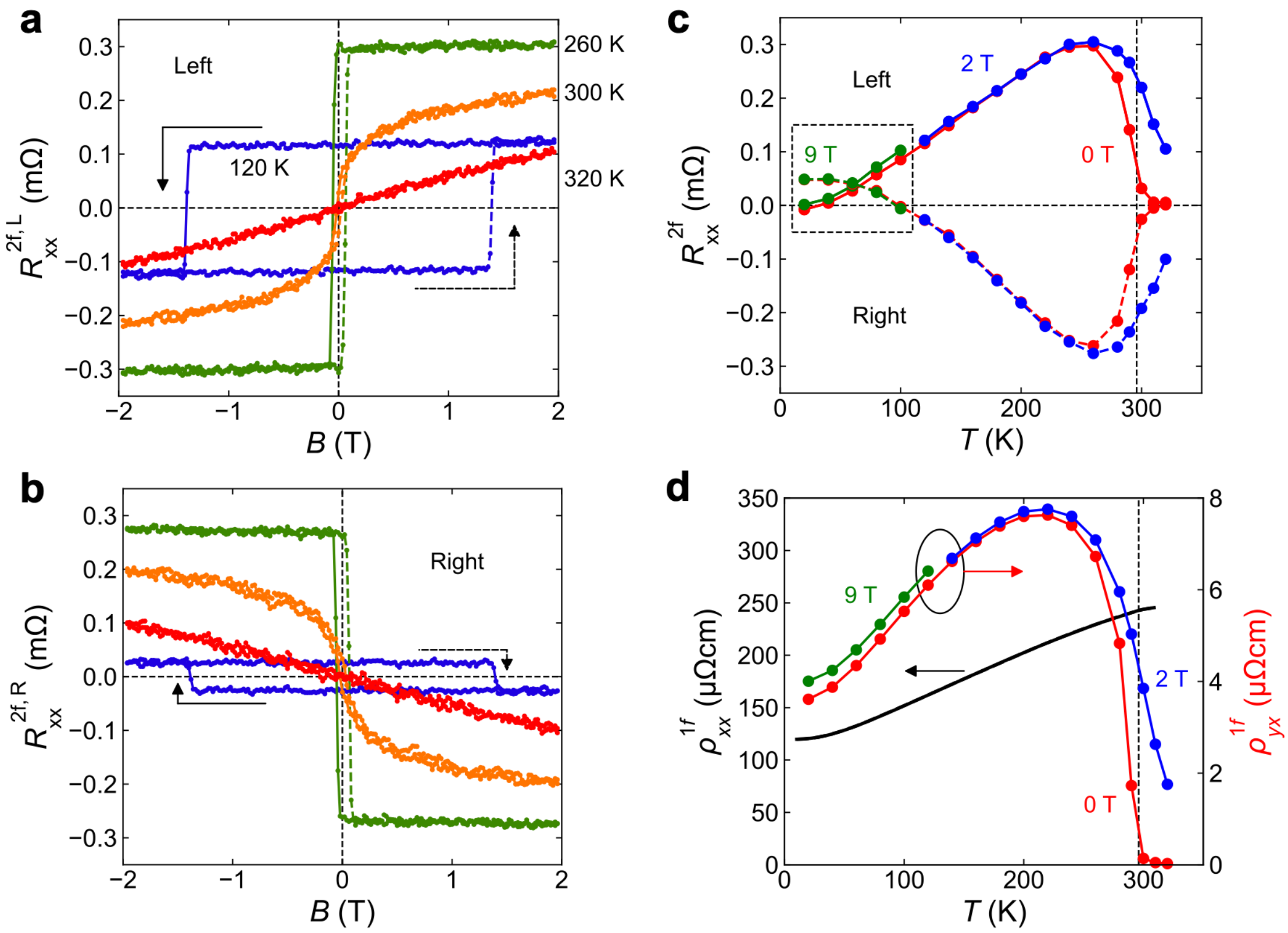

**Fig. 3: Magnetic-field and temperature dependences of the nonreciprocal resistance.** **a**,**b**, Magnetic-field dependence of the nonreciprocal resistance on the left (**a**) and right (**b**) sides of the FIB device at selected temperatures. Solid and dashed curves correspond to data taken in decreasing and increasing field, respectively, as indicated by the black arrows. The field is applied along the *c*-axis, and the ac electric current of $j = 8.3 \times 10^8$ $Am^{-2}$ at $f = 513$ Hz is applied. **c**, Temperature dependence of the nonreciprocal resistance on the left (solid) and right (dashed) sides. Boxed region: see main text. **d**, Temperature dependence of the first-harmonic longitudinal resistivity $\rho_{xx}^{1f}$ (black solid and dashed curves, which completely overlap for the two sides) and Hall resistivity $\rho_{yx}^{1f}$ (symbols and lines). Red, blue and green data in (**c**) and (**d**) correspond to $B$ = 0, +2 and +9 T, respectively. Vertical dashed lines in (**c**) and (**d**) indicate $T_C$.

**Magnetic switching in pristine $TbMn_6Sn_6$**

To separate the mechanism due to symmetry breaking from In substitution or a fixed device asymmetry, we measured an otherwise comparable microdevice made from pristine $TbMn_6Sn_6$. In the low-temperature easy-axis phase (Fig. 1b), opposing side surfaces show the same sign-reversed nonreciprocal response (Fig. 4a,b, Supplementary Fig. S7). Upon warming through the spin-reorientation transition near $T_{SR}$ = 315 K, $\boldsymbol{M}$ rotates into the kagome plane (Fig. 1c) and the longitudinal toroidal component $\boldsymbol{T} \cdot \boldsymbol{j} \sim (\boldsymbol{P} \times \boldsymbol{M}) \cdot \boldsymbol{j}$ switches off together with the side-antisymmetric signal (Fig. 4d). Above the $T_{SR}$, the side-antisymmetric signal reappears at above a threshold magnetic field $B_{SR}$, where the toroidal component $\boldsymbol{T} \cdot \boldsymbol{j}$ has a finite value (Fig. 4c), because an out-of-plane magnetic component is induced by the external magnetic field. There is a spin-fluctuation-assisted conspicuous enhancement at $B_{SR}$[26,27]. As such, spin-reorientation transition toggles boundary nonreciprocity without changing the crystal structure or device geometry. Moreover, since the parent pristine compound contains no substitutional In disorder, it provides a demonstration of the same phenomenon also in the chemically ordered centrosymmetric system. Together with the angular suppression in $TbMn_6Sn_{4.4}In_{1.6}$ (Supplementary Fig. S8), it directly tests the toroidal symmetry relation and extends the effect to room temperature.

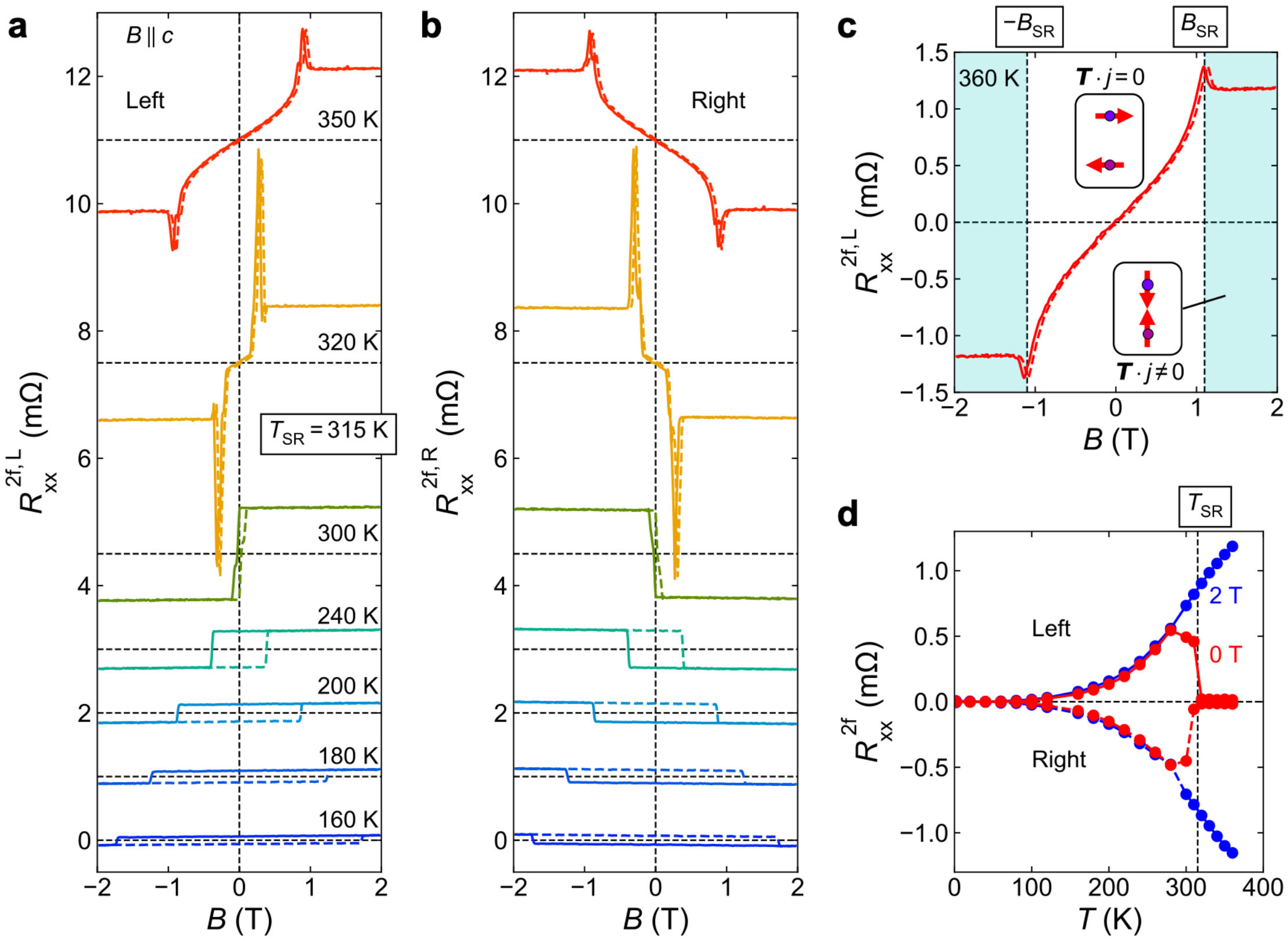

**Fig. 4: Magnetic switching of boundary nonreciprocity in pristine $TbMn_6Sn_6$.**

**a**,**b**, Magnetic-field dependence of the nonreciprocal resistance on the left (**a**) and right (**b**) sides of the FIB device of pristine $TbMn_6Sn_6$ at selected temperatures below and above the $T_{SR} \approx 315$ K. The data are shown with offsets as indicated by the horizontal black dashed lines. **c**, Magnetic-field dependence of the nonreciprocal resistance on the left side at 360 K. The vertical dashed lines in (**c**) indicate the magnetic-field induced spin-reorientation transition. Solid and dashed curves in (**a**-**c**) correspond to data taken in decreasing and increasing field, respectively. **d**, Temperature dependence of the nonreciprocal resistance on the left (solid) and right (dashed) sides in 0 T (red) and +2 T (blue). Vertical dashed line in (**d**) indicates the $T_{SR}$. The field is applied along the *c*-axis, and the ac electric current of $j = 8.3 \times 10^8$ Am$^{-2}$ at $f = 513$ Hz is applied.

## Discussion

Our measurements identify the polarity of a crystal surface as the inversion-breaking field that enables nonreciprocal transport in a centrosymmetric magnetic metal. Four observations establish this conclusion: the opposite signs at inversion-related side surfaces, the reversal with $\boldsymbol{M}$, the spatial potential map, and magnetic switching at the spin-reorientation transition of pristine $TbMn_6Sn_6$. The latter control is particularly restrictive because it changes the longitudinal component of $\boldsymbol{P} \times \boldsymbol{M}$ while preserving the material, device geometry and bulk centrosymmetric structure. The effect is therefore more than a second-harmonic signature of a particular doped compound; it is a boundary response selected by magnetic symmetry.

The materials implication is broad. In any centrosymmetric magnetic conductor, an exposed surface supplies a polar vector $\boldsymbol{P}$, and nonreciprocal longitudinal transport is symmetry-allowed wherever $(\boldsymbol{P} \times \boldsymbol{M}) \cdot \boldsymbol{j}$ is finite. Strong spin-orbit coupling, high mobility and topological surface bands can enhance the response, but they are not prerequisites for the underlying surface polarity. Conversely, the present transport measurements do not by themselves prove that Fermi arcs dominate the signal. Surface-resolved spectroscopy, electrostatic tuning or termination control will be needed to quantify that microscopic contribution. Treating the unavoidable crystal boundary as a functional symmetry element opens a route to zero-field rectification and magnetic switching across a wide range of conducting magnets, including centrosymmetric topological metals near and above room temperature.

## Methods

### Single-crystal growth and characterization

Single crystals of In-substituted $TbMn_6Sn_{4.4}In_{1.6}$ were grown in a high-power laser floating-zone furnace (LFZ furnace, Quantum Design). Polycrystalline feed and seed rods were first prepared. Tb, Mn, Sn and In were mixed in a molar ratio of 1:6:4.4:1.6 and arc-melted under G1-grade Ar gas atmosphere. The surface of the resulting polycrystal was ground to remove terbium oxide formed during melting. The material was then broken into pieces and remelted into rods of approximately 5 mm in diameter. The rods were installed in the LFZ furnace, translated at 1-2 $mmh^{-1}$ and counter-rotated at 20 r.p.m. The molten-zone temperature was monitored and maintained at approximately 800 °C by manual or feedback control of the laser power. Single-phase crystalline domains of approximately 1 × 0.5 × 0.2 $mm^3$ were obtained.

For comparison, pristine $TbMn_6Sn_6$ single crystals were grown by the Sn self-flux method[18]. Tb, Mn and Sn were mixed in a molar ratio of 1:6:30, placed in an alumina crucible and sealed in an evacuated quartz tube. The tube was heated to 1,100 °C over 10 h, held for 12 h and cooled to 500 °C at 6 $°Ch^{-1}$. At 500 °C, the excess flux was removed by centrifugation. Hexagonal-faceted single crystals of approximately 2 × 2 × 0.3 $mm^3$ were obtained.

Phase purity and single crystallinity were assessed by powder X-ray diffraction (RINT-TTR III, Rigaku) and Laue X-ray diffraction (m-Laue, Pulstec), respectively. The spatial homogeneity of In substitution was examined by energy-dispersive X-ray spectroscopy (XFlash 6-30, Bruker) in a scanning electron microscope (JSM-6701F, JEOL). Additional characterization is shown in Supplementary Fig. S1.

Magnetization was measured using a superconducting quantum interference device magnetometer (MPMS3, Quantum Design).

**Focused-ion-beam device fabrication**

Thin plates were cut from $TbMn_6Sn_{4.4}In_{1.6}$ and pristine $TbMn_6Sn_6$ single crystals by focused-ion-beam (FIB) milling (Helios 5UX, Thermo Fisher Scientific). Each plate was transferred with an in-situ micromanipulator (EasyLift) onto a sapphire substrate patterned with Au/Ti bilayer electrodes. Electrical contacts were formed by FIB-assisted Pt deposition, and the plates were milled into Hall-bar geometries. Device dimensions and measurement configurations are given in the corresponding figures and Supplementary information.

**Transport measurements**

Electrical transport was measured in a Physical Property Measurement System (PPMS, Quantum Design). First- and second-harmonic voltages were detected simultaneously using lock-in amplifiers (SR860, Stanford Research Systems), with the ac excitation supplied by the internal voltage output. The field-antisymmetric component was obtained from matched increasing- and decreasing-field branches. Representative raw data, harmonic phase information and the analysis procedure are provided in Supplementary Fig. S4.

## Acknowledgements

We thank X. Yu for support in FIB fabrication of microdevices; R-COMS_CEMS for technical support in the use of cleanroom facilities; D. Li, Y. Iwasa, M. Kawamura, T. Yokouchi and Y. Taguchi for fruitful discussions; and Y. Kaneko for designing the LFZ furnace.

## Author contributions

Y.T. conceived and supervised the project. D.Y. and H.O. grew and characterized the single crystals with assistance from A.K. M.T.B. and D.Y. fabricated the microdevices. D.Y. performed the transport measurements with assistance from M.T.B. and carried out the data analysis and visualization. D.Y. wrote the original draft. N.N. and Y.T. reviewed and edited the manuscript. All authors discussed the results and commented on the manuscript.

**Data and materials availability**

The data supporting the findings of this study are available within the Article.

**Funding**

This work was supported by JSPS KAKENHI Grant Nos. 23H05431 and 25K24552 (Y.T.), JSPS KAKENHI Grant Nos. 24H00197 and 24H02231 (N.N.), the RIKEN Junior Research Associate Program (D.Y.) and the RIKEN TRIP initiative (Fundamental Quantum Science Program, N.N.).

**Competing interests**

The authors declare no competing interests.

# Supplementary information:

# Crystal-surface polarity enables nonreciprocal transport in a centrosymmetric kagome magnet

Daiki Yamaguchi, Max T. Birch, Hibiki Onodera, Akiko Kikkawa,

Naoto Nagaosa, Yoshinori Tokura

**Table of contents:**

**Supplementary Figure S9:** Extended second-harmonic voltage mapping in a microdevice of $TbMn_6Sn_{4.4}In_{1.6}$ at 260 K

**Supplementary Figure S10:** Frequency dependence of boundary nonreciprocity at 260 K

**Supplementary Figure S11:** Dissipation of boundary nonreciprocity

**Supplementary Figure S12:** Device-width dependence of the nonreciprocal resistance

**Supplementary Note 1: Discussion on possible Joule-heating induced effects**

**A possible thermoelectric mechanism to explain the results provided in Fig. 2**

Since relatively large electric current density is applied during the nonreciprocal transport measurements, the Joule heating of a microdevice is potentially an extrinsic source of the observed second-harmonic voltage. The Joule heating power is proportional to the square of the applied electric current as $RI^2$. Here, $R$ is determined by the contact resistance at the Pt-deposition, between the FIB fabricated material and the conductive Au pattern on the insulating sapphire substrate. The contact resistance is typically at around a hundred ohms and dominates the Joule heating effect since the resistance of the FIB fabricated material itself is smaller by several orders of magnitude than the contact resistance. Therefore, we may assume the heat flows from both of the source/drain ($I$+/$I$-) ac current input electrodes at the ends of the microdevice.

Such an effect of the Joule heating created by the ac electric current is separated into two components: dc and ac. When the ac current $I = I_0\sin(\omega t)$ is applied, the Joule heating power is,

$$RI^2 = RI_0^2 \sin^2(\omega t) = \frac{RI_0^2}{2} - \frac{RI_0^2}{2}\cos(2\omega t)\,. \tag{S1}$$

The first term is the dc component, and the second term is the ac second-harmonic out-of-phase component. The former dc component only increases the device temperature from the base temperature in equivalence, and the lock-in measurement does not detect this dc signal induced effects. On the other hand, the latter ac component could be detected through the thermoelectric (Seebeck or Nernst) effects in the second-harmonic lock-in measurement, only if the ac component of the heating is harmonic to the input ac electric current input.

Thus, the longitudinal (Seebeck) and transverse (Nernst) thermoelectric voltages appear as an ac second-harmonic out-of-phase component (detectable in the $2f$ lock-in measurement) as follows[1]. The Seebeck voltage is,

$$V_{\mathrm{x}} = S_{\mathrm{xx}} \cdot \partial_{\mathrm{x}} T \cdot l, \tag{S2}$$

where $S_{\mathrm{xx}}$ is the Seebeck thermopower, $l$ is the distance between the voltage terminals, and $\partial_{\mathrm{x}} T = -\frac{\Delta T}{l} = -\frac{T_{\mathrm{hot}} - T_{\mathrm{cold}}}{l}$ is the thermal gradient. Note that the $V_{\mathrm{x}}$ is measured from the cold side of the device. The Nernst voltage is,

$$V_{\mathrm{y}} = -S_{\mathrm{xy}} \cdot \partial_{\mathrm{x}} T \cdot w, \tag{S3}$$

where $S_{\mathrm{xy}}$ is the Nernst thermopower, and $w$ is the distance between the transverse voltage terminals (width of the microdevice). In the present case, the thermal gradient is proportional to the Joule heating $RI^2$. Since both $V_{\mathrm{x}}$ and $V_{\mathrm{y}}$ are proportional to the thermal gradient, these could be observed as ac second-harmonic out-of-phase component in the lock-in measurement.

To explain the results provided in main text Fig. 2 in terms of thermoelectric effects, we must assume the equal application of heater powers from the two source/drain electrodes at both ends of the device. The ac heat goes into the device along the longitudinal direction. At this moment, ac thermal gradients of opposite sign are formed at around terminals 1 and 3 (see main text Fig. 2d). At the terminal 2, in the middle, where the two thermal gradient merge, thermal gradient is zero. Under the magnetic field perpendicular to the device plane, Nernst effect creates voltage drops at $V_{\mathrm{y}}^1$ and $V_{\mathrm{y}}^3$, in opposite directions with the same magnitude, while the voltage drop is zero in the middle at $V_{\mathrm{y}}^2$, since the thermal gradient at the middle is cancelled out. Until here, the thermoelectric scenario would be consistent with the second-harmonic potential mapping reported in main text Fig. 2. Here, let us quantitatively estimate how large the ac thermal gradient formed along the microdevice in the case of main text Fig. 2 is. From Fig. 2h,i, $-V_y^{2f,1} = V_y^{2f,3} \approx 500$ nV. Using $S_{xy} \approx -1.5$ μVK$^{-1}$, $w = 782$ nm, from equation S3, $-\partial_x T \approx 0.4$ Kμm$^{-1}$.

To further validate the above scenario, the following issues should be addressed.

1. What is the distribution of ac thermal gradient along the whole microdevice?
2. Is the Seebeck effect induced signal consistent with the Nernst effect induced signal?
3. How does the formation of the ac thermal gradient change with frequency?
4. Is the estimated magnitude of the ac thermal gradient reasonable?

To consider these points, a longer microdevice with increased number of voltage terminals (shown in Supplementary Fig. S2c) was prepared, enabling extended potential mapping along the device. Frequency dependence of the second harmonic response was also measured in the device. In conclusion, the above thermoelectric scenario cannot explain the observed phenomena as follows.

**Extended potential mapping along the microdevice hinders thermoelectric scenario**

The extended mapping of the voltage potential is shown in Supplementary Fig. S9. First of all, the first harmonic longitudinal voltage distribution is perfectly linear along the microdevice (Fig. S9b). As a result, the first harmonic Hall effect is also homogeneous along the device (Fig. S9c). Therefore, the dc (not ac) thermal gradient from the Joule heating is not severe in the device, otherwise, the resistance near the source/drain electrodes would be higher compared to the middle of the device. Once the dc thermal gradient is confirmed to be negligible, let us discuss the ac thermal gradient potentially detected in the second-harmonic voltage measurements. Figures S9d and S9e show the second-harmonic potential mapping of field symmetric and anti-symmetric components, respectively. The potential distribution of the field antisymmetric component in Fig. S9e is approximately linear with position along the device. If the ac thermal gradient is strongly localized near the source/drain electrodes, the potential at terminal 1 and 7 should be particularly enhanced, but it is not the case here. As a result of the opposite-sign longitudinal potential distribution on side surfaces, a corresponding transverse potential $V_y^{2f}$ distribution is also measured. It is shown in Fig. S9f, showing zero signal at terminal 4 in the middle. If these signals were originated from the Nernst effect, the created ac thermal gradient would be zero in the middle and almost linearly increases toward the source/drain electrodes.

Now, let us check whether the profile of ac thermal gradient described above induces consistent Seebeck effect induced signal. The Seebeck effect induced signal is observed as a field-symmetric component of the second harmonic voltage. The data is shown in Fig. S9d. The potential mapping is strongly non-monotonic and asymmetric against the position. Between terminals 1-2, 2-3, 3-4, there are only small signal. On the other hand, between 4-5, 5-6, 6-7, there are larger signals of around 5-10 μV. This means the ac thermal gradient distribution is asymmetric along the device, while this contradicts the conclusion obtained from the Nernst effect induced signal, discussed above. In other words, if the field-symmetric component is attributed to the Seebeck effect, it does not behave in a way that would be consistent with the field-anti-symmetric signal (focus of this work). As such, extended second harmonic potential mapping reveals that the appropriate model of the distribution of ac thermal gradient to explain all the results without contradiction is impossible to be constructed.

**Frequency independent second-harmonic response up to 10 kHz**

Another aspect we should consider is whether the ac thermal gradient as a result of the ac component in equation S1 is harmonic to the frequency of the input ac electric current. We need to consider the thermal diffusion through the FIB device, which is substantially distinct from the case of the application of electric current. Heat propagation into the FIB device, particularly at the low-conductance interface between the Pt-deposition and the FIB fabricated material surface is expected to be highly dissipative or the thermal wavelength is extremely short. Moreover, the charge and entropy relaxations are distinct in scattering mechanism. In terms of the electronic transport, the inelastic electron-phonon scattering also contributes to entropy relaxation, while not in charge relaxation. The phonon contribution to the entropy relaxation is also included only in the entropy relaxation. Considering such a difference in transport mechanisms, the synchronized conversion from the applied ac electric current and the resulting ac thermal gradient should not be straightforwardly assumed.

The data shown in Supplementary Fig. S10 experimentally demonstrate the frequency independence up to around 10 kHz (Fig. S10a,b). This trend is irrelevant to the terminal pair combinations for the voltage measurement, i.e., completely scalable frequency dependence across the device, (Fig. S10c). The reduction in the magnitude of the response at around 10 kHz is seemingly an upper limit for the proper ac excitation input to a microdevice, rather than the cutoff frequency of the thermal diffusion.

**Non-trivial heat propagation around the FIB device**

The possible heat sinks for the generated Joule heating are, sapphire substrate (including the patterned Au), helium gas atmosphere in the PPMS chamber, FIB fabricated material itself, and so on. Given the larger thermal conductivities of gold and sapphire and the larger volume of the substrate compared to the FIB fabricated material, most of the heat is expected to dissipate into the substrate, rather than the device itself. Also, some of the heat may go into the chamber atmosphere. Note again that the heat sources are at both ends of the microdevice at the source/drain electrodes, i.e., the material is in between the two different heat sources, which usually have different contact resistance. The possible heat sink around the device body is the helium gas atmosphere since there is a space beneath the material, which thermally decouples the substrate surface and the material. The voltage terminals on the sides may also work as heat sinks. Such a complex, non-trivial environment around the microdevice generally results in asymmetric heat propagation through the microdevice, which is inconsistent with the present results.

## Supplementary Note 2: Nonreciprocal boundary resistance as a probe of surface relaxation time

The surface nonreciprocal resistance can be related to a surface relaxation time as follows. Consider surface transport in the Boltzmann transport theory, including the leading nonlinear term in the electric field. In this regime, $\sigma_s^{(1)} \propto \tau_s$ and $\sigma_s^{(2)} \propto \tau_s^2$, where $\sigma_s^{(1)}$ and $\sigma_s^{(2)}$ are the surface first- and second- order conductivities, respectively, and $\tau_s$ is the surface relaxation time[2]. The surface current density can then be written phenomenologically as,

$$j_s = \sigma_s^{(1)} E_s + \sigma_s^{(2)} E_s^2 + \cdots, \tag{S4}$$

where $j_s$ is the surface current density and $E_s$ is the surface electric field. Conversely,

$$E_s = \rho_s^{(1)} j_s + \rho_s^{(2)} j_s^2 + \cdots. \tag{S5}$$

where $\rho_s^{(1)}$ and $\rho_s^{(2)}$ are the surface first- and second-order resistivities, respectively. Substituting equation S4 into equation S5 and retaining terms through second order in $E_s$

gives

$$E_{\mathrm{s}} = \rho_{\mathrm{s}}^{(1)}\sigma_{\mathrm{s}}^{(1)}E_{\mathrm{s}} + \left(\rho_{\mathrm{s}}^{(1)}\sigma_{\mathrm{s}}^{(2)} + \rho_{\mathrm{s}}^{(2)}{\sigma_{\mathrm{s}}^{(1)}}^{2}\right)E_{\mathrm{s}}^{2}. \tag{S6}$$

Comparison of the coefficients of $E_{\mathrm{s}}$ and $E_{\mathrm{s}}^2$ yields

$$1 = \rho_{\mathrm{s}}^{(1)}\sigma_{\mathrm{s}}^{(1)}, \tag{S7}$$

$$0 = \rho_{\mathrm{s}}^{(1)}\sigma_{\mathrm{s}}^{(2)} + \rho_{\mathrm{s}}^{(2)}{\sigma_{\mathrm{s}}^{(1)}}^{2}. \tag{S8}$$

Equation S7 gives the usual relation between the linear resistivity and conductivity,

$$\rho_{\mathrm{s}}^{(1)} = \frac{1}{\sigma_{\mathrm{s}}^{(1)}}. \tag{S9}$$

and equation S8 gives

$$\rho_{\mathrm{s}}^{(2)} = -\frac{\rho_{\mathrm{s}}^{(1)}\sigma_{\mathrm{s}}^{(2)}}{{\sigma_{\mathrm{s}}^{(1)}}^{2}} = -\frac{\sigma_{\mathrm{s}}^{(2)}}{{\sigma_{\mathrm{s}}^{(1)}}^{3}}. \tag{S10}$$

Because $\sigma_{\mathrm{s}}^{(1)} \propto \tau_{\mathrm{s}}$ and $\sigma_{\mathrm{s}}^{(2)} \propto \tau_{\mathrm{s}}^2$, both $\rho_{\mathrm{s}}^{(1)}$ and $\rho_{\mathrm{s}}^{(2)}$ scale as $\tau_{\mathrm{s}}^{-1}$. Although $\rho_{\mathrm{s}}^{(1)}$ is difficult to isolate experimentally, $\rho_{\mathrm{s}}^{(2)}$ is accessible through the present nonreciprocal transport measurement. The same argument applies for the bulk transport, i.e., $\rho_{\mathrm{b}}^{(2)} \propto \tau_{\mathrm{b}}^{-1}$, while it is negligible because of the bulk centrosymmetry. This identification is only qualitative because $\rho_{\mathrm{b}}^{(1)}$ and $\rho_{\mathrm{s}}^{(1)}$ are generally different.

In the present case of a Weyl semimetal, the device body, not only its surface, is also conductive. Therefore, the surface current is expected to dissipate into the bulk. However, the $2f$ boundary response still shows a distinct temperature dependence compared to the $1f$ transport as discussed above. Here, we provide a result of the measurement in different configuration to the other measurements in the manuscript (Supplementary Fig. S11) to further capture the nature of the surface transport. As shown in Fig. S11a, the ac electric current is applied using the previously voltage reading terminals at 7 (see Fig. S9a). The 1*f* and 2*f* longitudinal voltages are measured at each remaining pair of voltage terminals. The observed signal is plotted against the relative position along the device in Fig. S11b and S11c. The finite and opposite sign (between upper and lower sides) 1*f* voltage is only observed at the terminal pairs 5-6, which are just beside the ac current flow. It is vanishingly small at other pairs (Fig. S11b). The bulk transport seems to be dominantly detected here. In contrast, the small but finite 2*f* voltage is still observed at terminal pairs 4-5 not only at 5-6 (Fig. S11c). This difference in the decay between 1*f* and 2*f* signals is additional evidence consistent with the existence of dissipative surface current, of which the relaxation time is distinct from the bulk transport. Moreover, the delayed decay in the 2*f* signal (i.e., the observation of finite signal even at 4-5) is hard to be understood by the thermoelectric scenario described in Supplementary Note 1, since the position of the heat-sources are different from the case in the previous discussion.

**Supplementary Note 3: Device-width dependence**

We compared devices with widths $w$ = 396, 782 and 1,780 nm and otherwise nominally similar dimensions (#1-3, in Fig. S2a). The same current density in bulk $j_b$ was used in each device. Based on the discussion in Supplementary Note 2, we use $R_{xx}^{2f} = V_x^{2f}/I$ as an empirical proxy for changes in $\tau_s^{-1}$. All three devices reproduce the enhancement towards $T_C$, and the low-temperature sign change, establishing reproducibility across independently fabricated devices (Supplementary Fig. S12a). The two wider devices show similar nonreciprocal resistances over an intermediate temperature range, whereas the narrowest device is systematically suppressed (Supplementary Fig. S12b). Near $T_C$, the magnitude increases with width.

These data reveal a crossover when the separation between opposing surfaces approaches several hundred nanometres. Because a measured side voltage in a metallic Hall bar reflects both a nonlinear boundary current and its redistribution through the conducting bulk, the width dependence cannot be converted directly into a surface-state localization length. However, since the surface conductance is experimentally shown to be negligibly small compared to the bulk conductance, $j_b$ may be approximated to be constant in all three devices. Therefore, the surface current density is approximately proportional to the applied current $I$, rather than $j_b$. Based on this approximation, the comparison of the $R_{xx}^{2f} = V_x^{2f}/I$ is still justified, and the suppression in the narrowest device is possibly related to equilibration or coupling between opposing boundaries. Possible pathways include bulk-mediated propagation, a Weyl orbit linking surface states, or propagation around the end surfaces near the current electrodes. The width series therefore supports a boundary-sensitive origin, while a quantitative microscopic assignment will require modelling of the full device current distribution, which is highly non-trivial in such a conducting material.

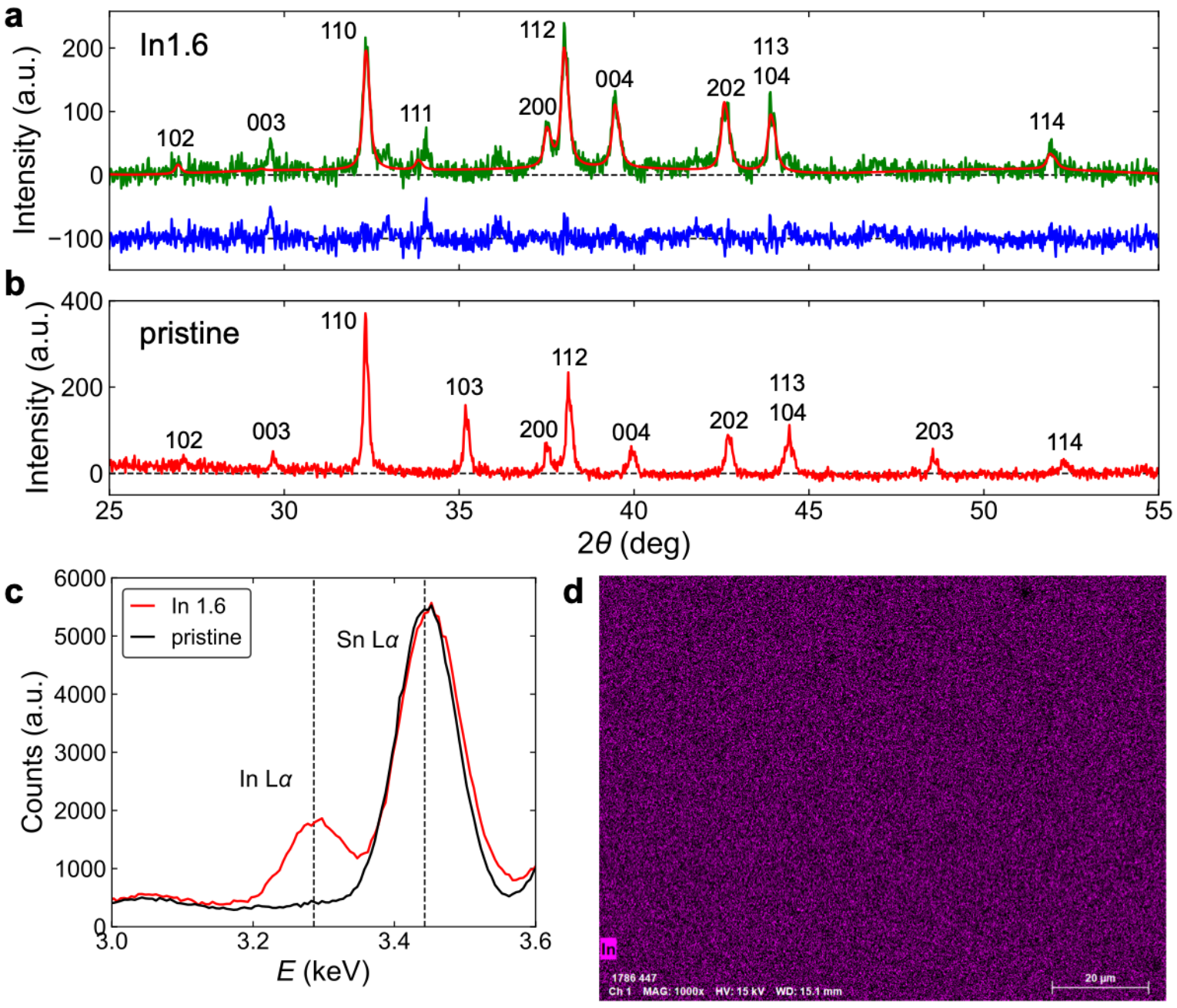

**Supplementary Fig. S1: X-ray and compositional characterization of $TbMn_6Sn_{4.4}In_{1.6}$ and $TbMn_6Sn_6$ single crystals.**

**a**, **b**, Powder X-ray diffraction patterns of $TbMn_6Sn_{4.4}In_{1.6}$ (**a**) and $TbMn_6Sn_6$ (**b**). Peaks are labelled by their Miller indices. In (**a**), green and red curves show the measured and calculated intensities, respectively, and the blue curve is their difference. In (**b**), the red curve shows the measured intensity. Both samples are in single phase. The differences between the peak patterns reflect expansion of the *c*-axis lattice parameter due to partial substitution of In for Sn. The calculated lattice constants are: $a$ = 0.5532(2) nm, $c$ = 0.9127(3) nm in $TbMn_6Sn_{4.4}In_{1.6}$, and $a$ = 0.55286(1) nm, $c$ = 0.90179(2) nm in $TbMn_6Sn_6$. **c**, Energy-dispersive X-ray spectra of the In-substituted and pristine crystals, normalized to the Sn L$\alpha$ peak. The In L$\alpha$ peak is present only in the In-substituted crystal. **d**, Elemental map of In at the surface of the In-substituted crystal, showing a homogeneous spatial distribution.

**a**

| | $L$ (μm) | $l$ (μm) | $w$ (nm) | $t$ (μm) | Figs |
|---|---|---|---|---|---|
| In1.6 #1 | 12 | 7.6 | 396 | 1.26 | 1, 3, S6, S8, S12 |
| In1.6 #2 | 12 | 7.6 | 782 | 1.26 | 2, 3d, S3, S4, S5, S12 |
| In1.6 #3 | 12 | 7.6 | 1780 | 1.26 | S12 |
| In1.6 #4 | 22 | 18 | 1890 | 1.13 | S9, S10, S11 |
| pristine #5 | 11 | 7.8 | 850 | 0.856 | 1, 4, S7 |

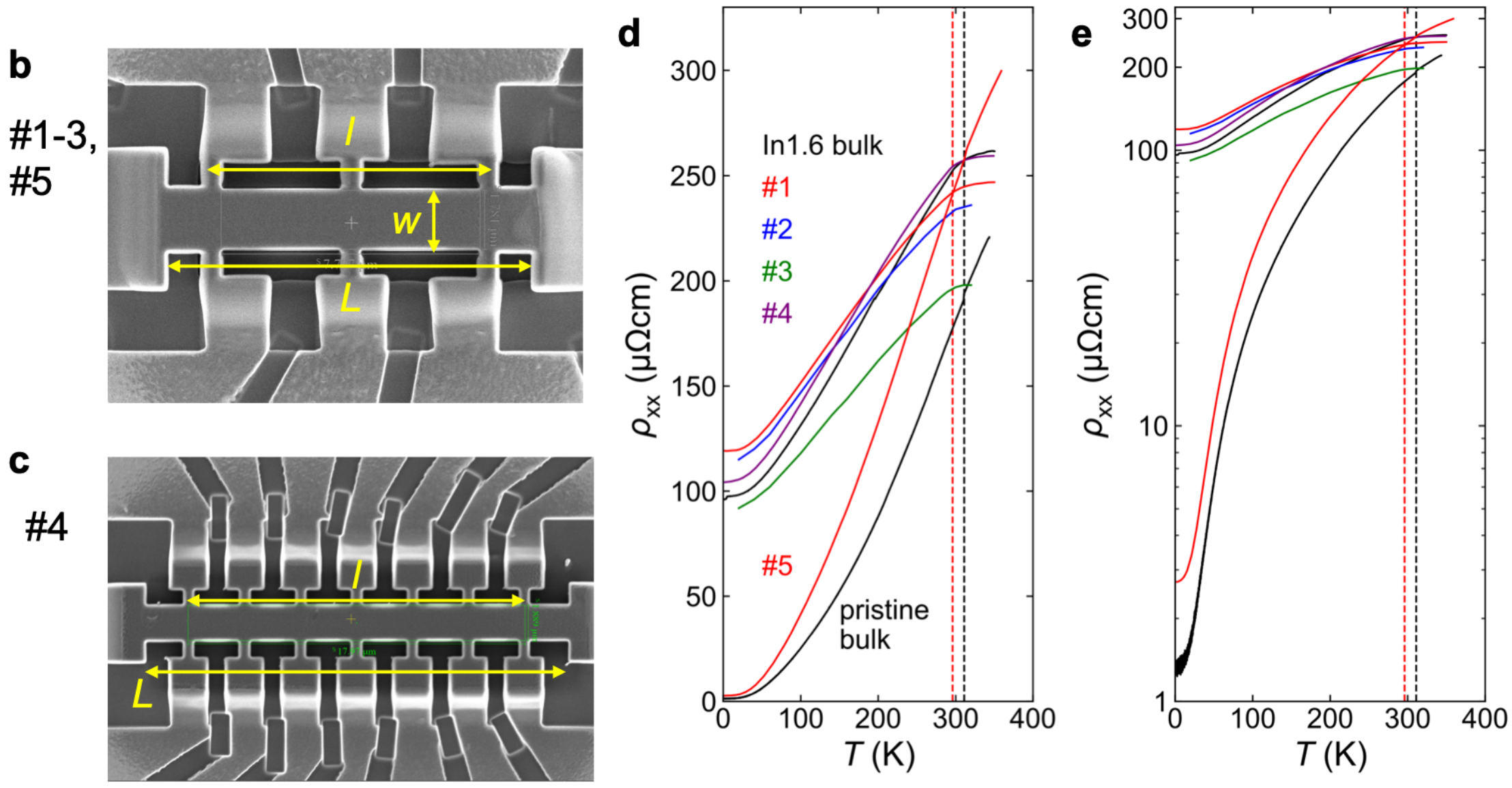

**Supplementary Fig. S2: List of FIB-fabricated microdevices and $\rho$-$T$ profiles.**

**a**, List of FIB devices and typical dimensions. $t$ is the thickness of the devices. The corresponding figure numbers indicate the figure where the data measured on the device is shown. **b**, **c**, SEM images of the FIB devices: Devices #1-3 and #5 are in the shape of (**b**), while #4 is a device in (**c**) with increased number of electrodes. **d**, Comparison of temperature dependence of the longitudinal resistivity among all the devices in (**a**) and the single crystals from which the FIB devices are fabricated. **e**, Vertical axis of (**d**) is changed into the logarithmic scale to focus on the difference in residual resistivity. Red and black vertical dashed lines in (**d**) and (**e**) indicate the $T_C$ and the $T_{SR}$, respectively. Variation in absolute values may partly come from the measurement error of the device geometry.

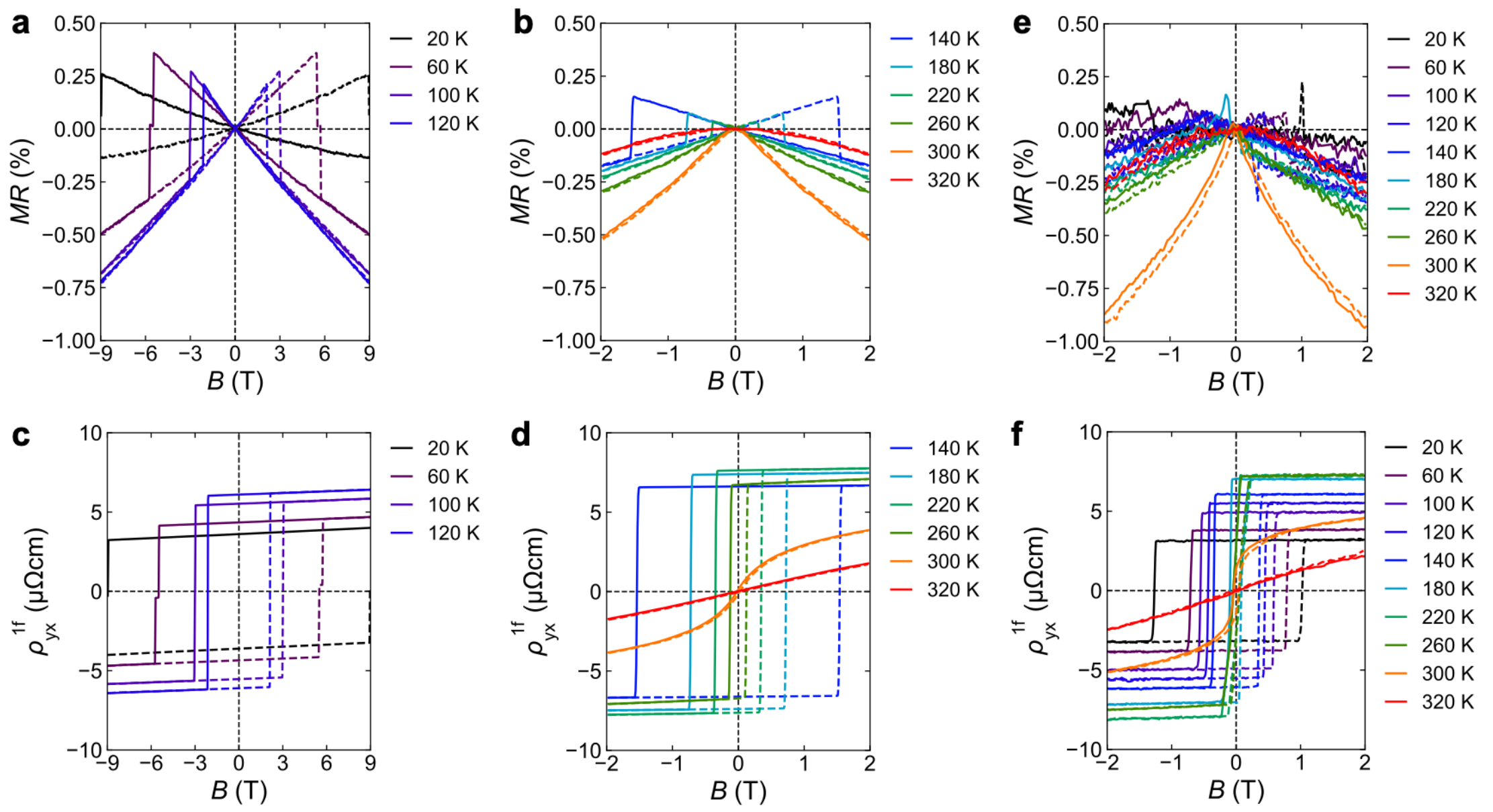

**Supplementary Fig. S3: Magnetic transport properties of a FIB fabricated microdevice and a bulk single crystal of $TbMn_6Sn_{4.4}In_{1.6}$.**

**a**, **b**, **e**, Magnetic-field dependence of the magnetoresistance in the FIB fabricated microdevice (**a**, **b**) and bulk crystal (**e**). **c**, **d**, **f**, Magnetic-field dependence of the Hall resistivity in the FIB device (**c**, **d**) and bulk crystal (**f**). Solid and dashed curves correspond to data taken in decreasing and increasing field, respectively. The magnitudes of the magnetoresistance and Hall resistivity are comparable between the device and bulk crystal. The coercive field is substantially larger in the FIB device, possibly owing to enhanced domain-wall pinning. An asymmetric hysteresis loop (exchange-bias-like behaviour) is observed only in the bulk crystal.

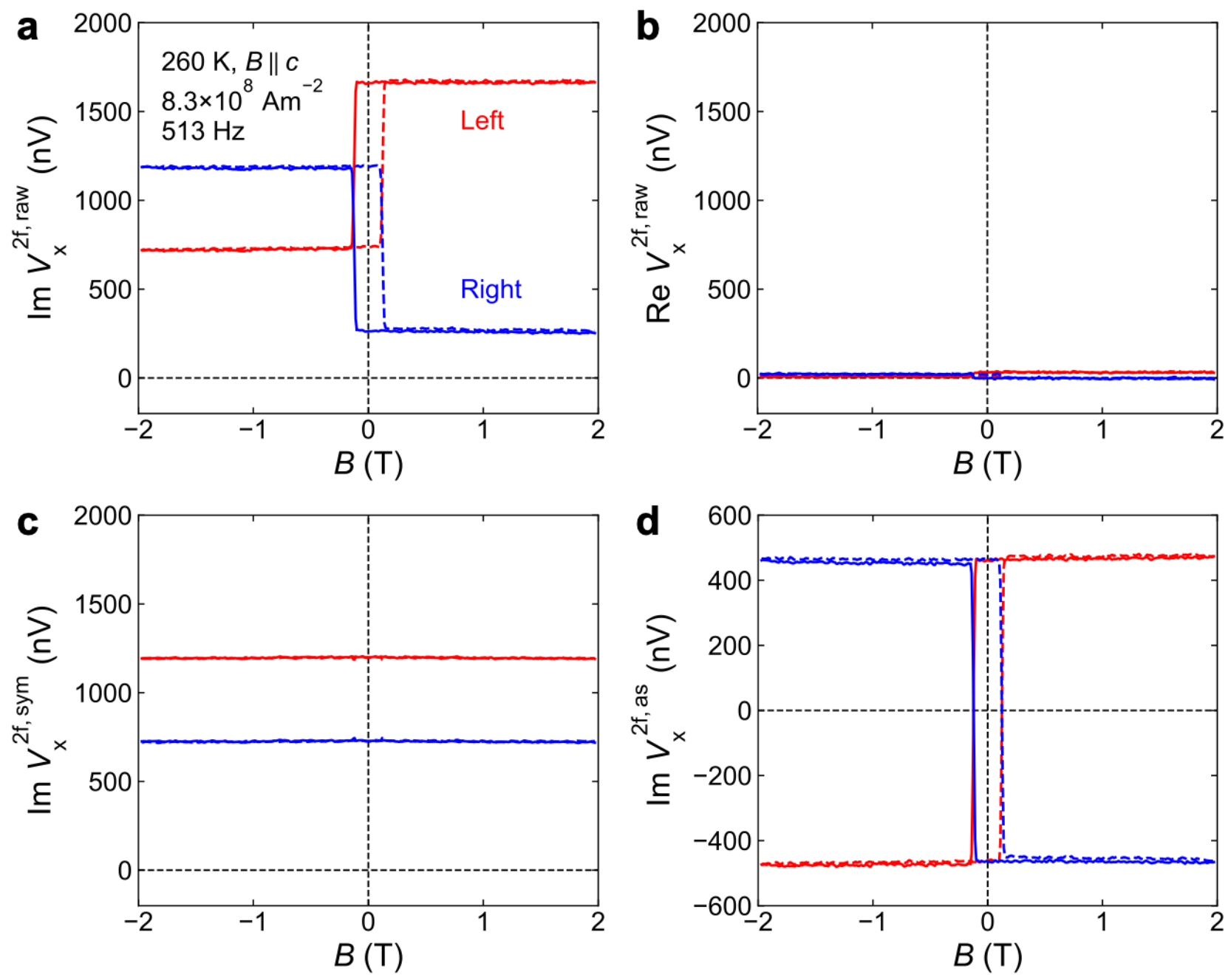

**Supplementary Fig. S4: Representative raw data and analysis of the nonreciprocal second-harmonic voltage in $TbMn_6Sn_{4.4}In_{1.6}$.**

**a**, **b**, Magnetic-field ($B \parallel c$) dependence of the imaginary (**a**) and real (**b**) components of the second-harmonic ($2f$) longitudinal voltage at 260 K. An ac current density $j = 8.3 \times 10^8$ Am$^{-2}$ at $f = 513$ Hz is applied. Solid and dashed curves correspond to data taken in decreasing and increasing field, respectively. The real component is negligible compared with the imaginary component. **c**, **d**, Field-symmetric (**c**) and field-antisymmetric (**d**) components of the imaginary signal in (**a**). The symmetric component may arise from device heating, whereas the antisymmetric component is defined as the nonreciprocal $2f$ voltage in the main text and is consistent with the symmetry analysis. The symmetric components differ in magnitude between the left and right sides, whereas the antisymmetric components have comparable magnitudes, indicating distinct origins.

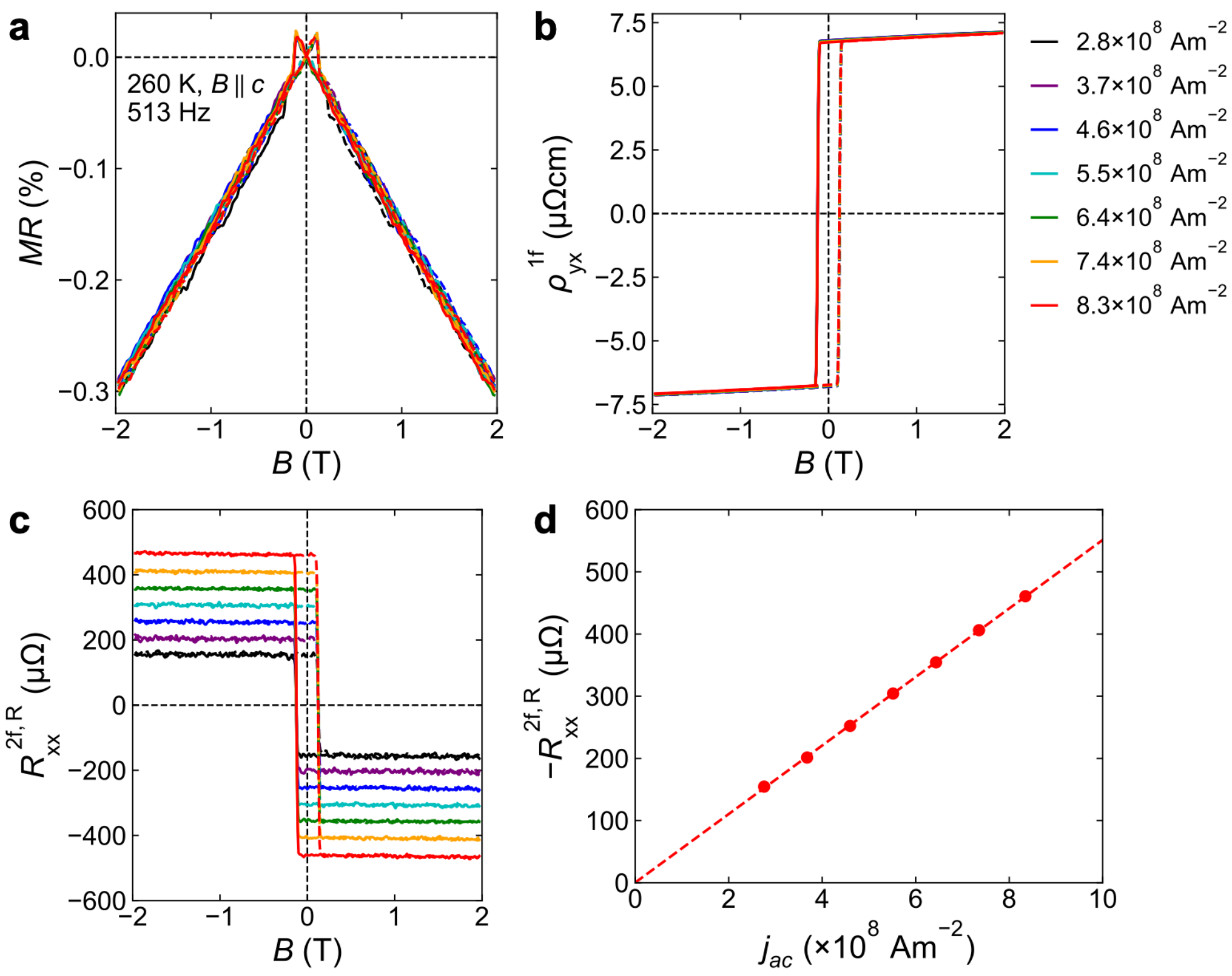

**Supplementary Fig. S5: Current-density dependence of the transport properties in $TbMn_6Sn_{4.4}In_{1.6}$.**

**a**-**c**, Magnetic-field dependence of the magnetoresistance (**a**), Hall resistivity (**b**) and nonreciprocal resistance (**c**) on the right side at 260 K for several current densities. The magnetoresistance and Hall resistivity are independent of current density. Solid and dashed curves correspond to data taken in decreasing and increasing field, respectively. **d**, Current-density dependence of the nonreciprocal resistance extracted from (**c**), confirming a linear dependence on current density.

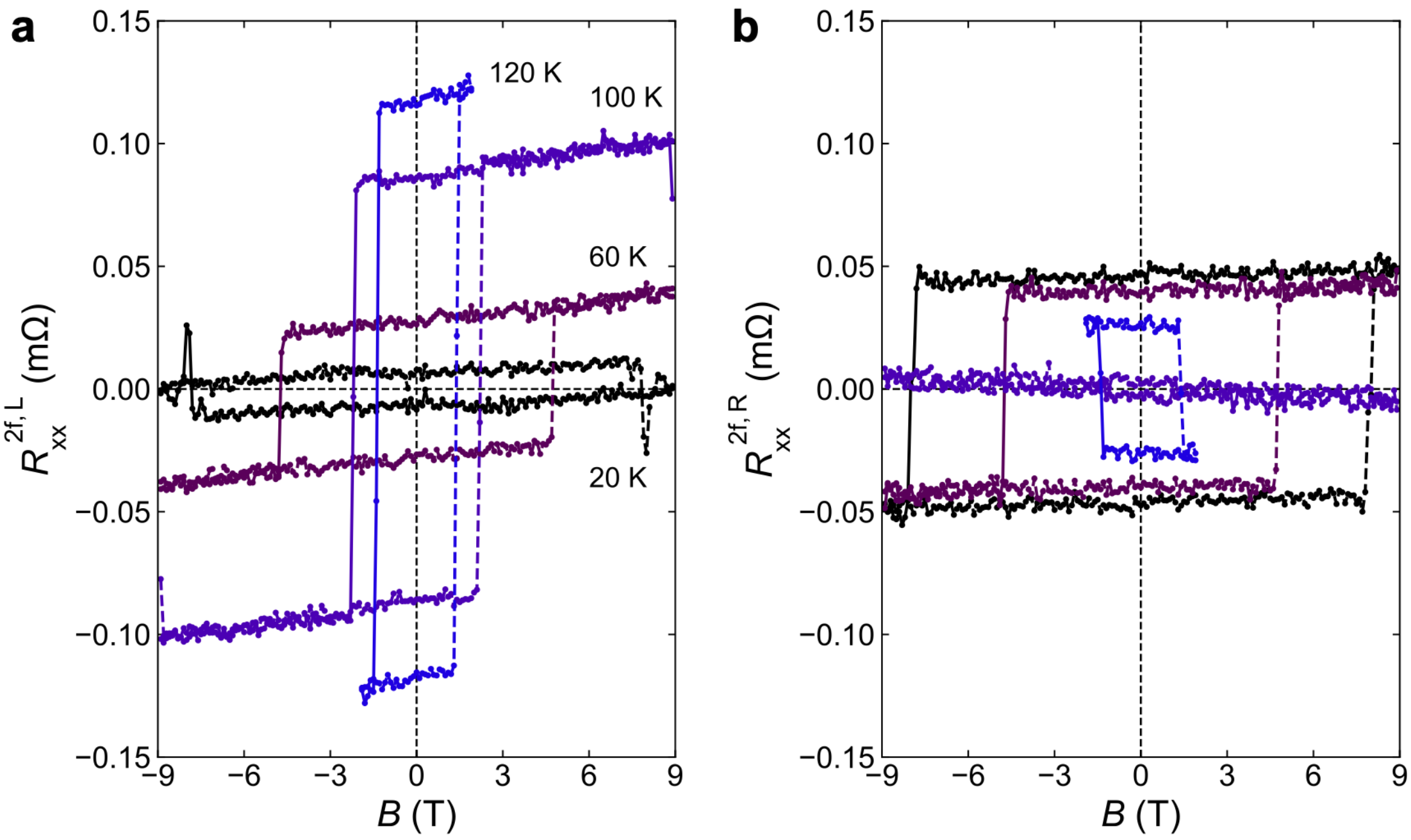

**Supplementary Fig. S6: Low-temperature magnetic-field dependence of the nonreciprocal resistance in $TbMn_6Sn_{4.4}In_{1.6}$.**

**a**, **b**, Magnetic-field dependence up to ±9 T of the nonreciprocal resistance on the left (**a**) and right (**b**) sides of the FIB device at low temperatures. The field is applied along the *c*-axis, and the ac current density $j = 8.3 \times 10^8$ Am$^{-2}$ at $f = 513$ Hz is applied. Solid and dashed curves correspond to data taken in decreasing and increasing field, respectively. A temperature-driven sign reversal is observed.

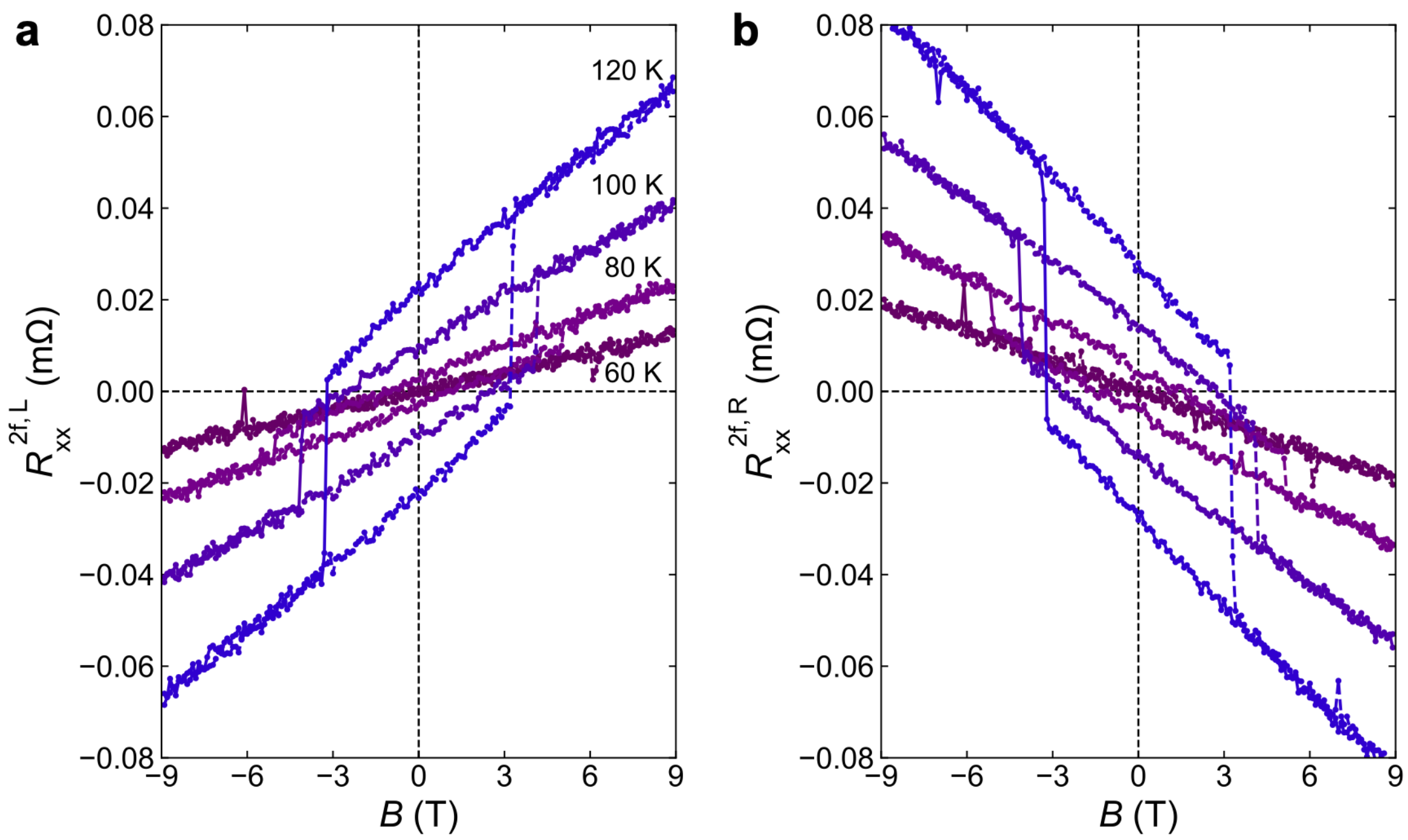

**Supplementary Fig. S7: Low-temperature magnetic-field dependence of the nonreciprocal resistance in $TbMn_6Sn_6$.**

**a**, **b**, Magnetic-field dependence up to ±9 T of the nonreciprocal resistance on the left (**a**) and right (**b**) sides of the FIB device at low temperatures. The field is applied along the *c*-axis, and the ac current density $j = 8.3 \times 10^8$ Am$^{-2}$ at $f = 513$ Hz is applied. Solid and dashed curves correspond to data taken in decreasing and increasing field, respectively. In contrast to the case in $TbMn_6Sn_{4.4}In_{1.6}$, the spontaneous signal in zero field is vanishingly small below 60 K, reflecting the lower residual resistivity of the material. The field-linear component is enhanced in the pristine $TbMn_6Sn_6$ compared to $TbMn_6Sn_{4.4}In_{1.6}$.

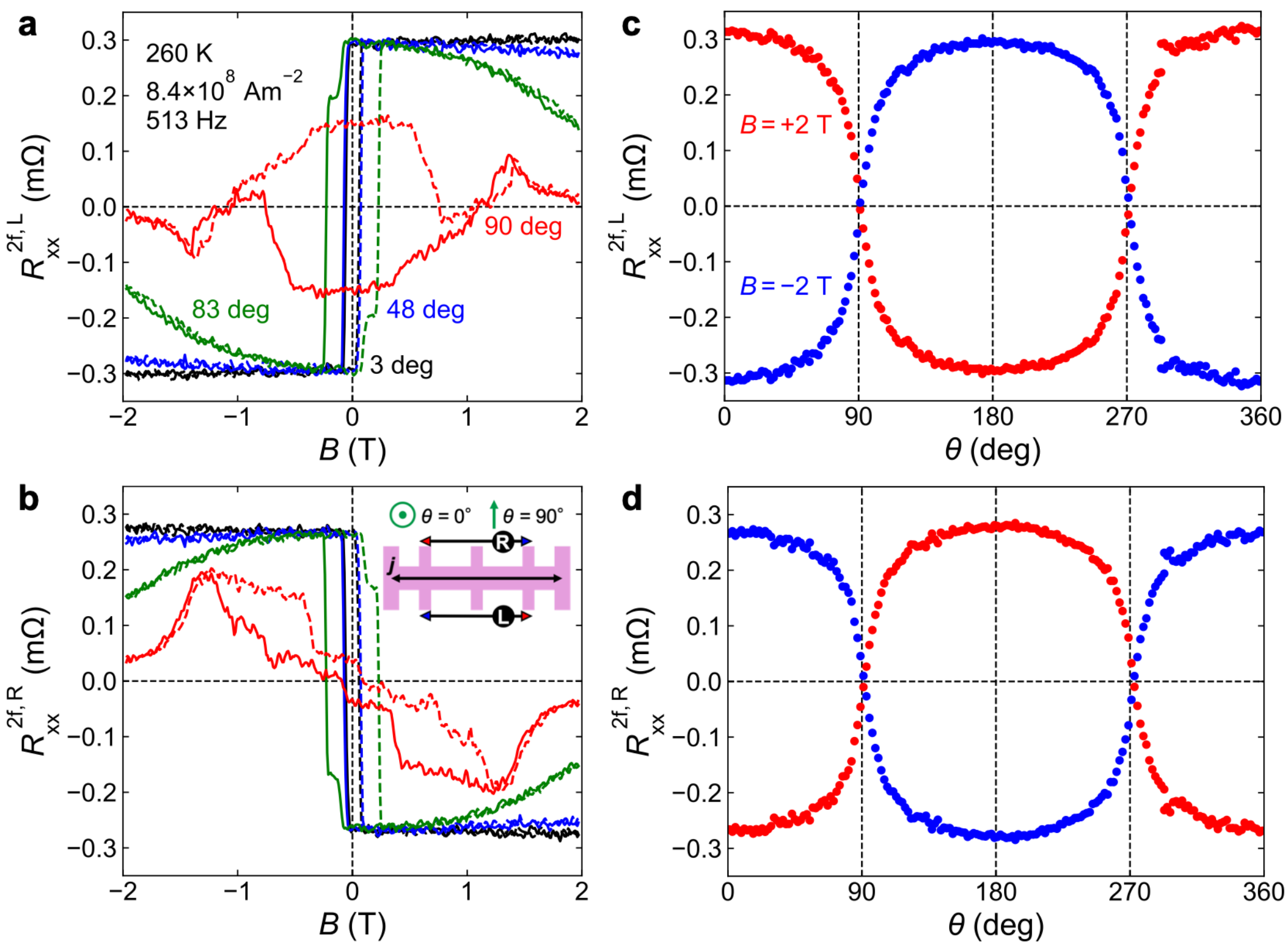

**Supplementary Fig. S8: Magnetic-field-angle dependence of the nonreciprocal resistance at 260 K in $TbMn_6Sn_{4.4}In_{1.6}$.**

**a**, **b**, Magnetic-field dependence of the nonreciprocal resistance on the left (**a**) and right (**b**) side surfaces of the device for field angles $\theta$ = 3°, 48°, 83° and 90° from the *c*-axis (black, blue, green and red curves, respectively). Solid and dashed curves correspond to data taken in decreasing and increasing field. The inset in (**b**) defines the rotation direction. When the magnetic field is scanned at $\theta$ = 90°, there are hysteretic (different shape on each side) signals at low field. This may reflect the complex domain switching process at each surface. Substantially, the signal at high field is suppressed to be zero at $\theta$ = 90°, consistent with zero toroidal moment. **c**, **d**, Angular dependence of the nonreciprocal resistance on the left (**c**) and right (**d**) side surfaces at $B$ = +2 T (red) and −2 T (blue). The response vanishes for $\boldsymbol{P} \parallel \boldsymbol{M}$ ($\theta$ = 90° and 270°), for which $\boldsymbol{T} \propto \boldsymbol{P} \times \boldsymbol{M} = 0$.

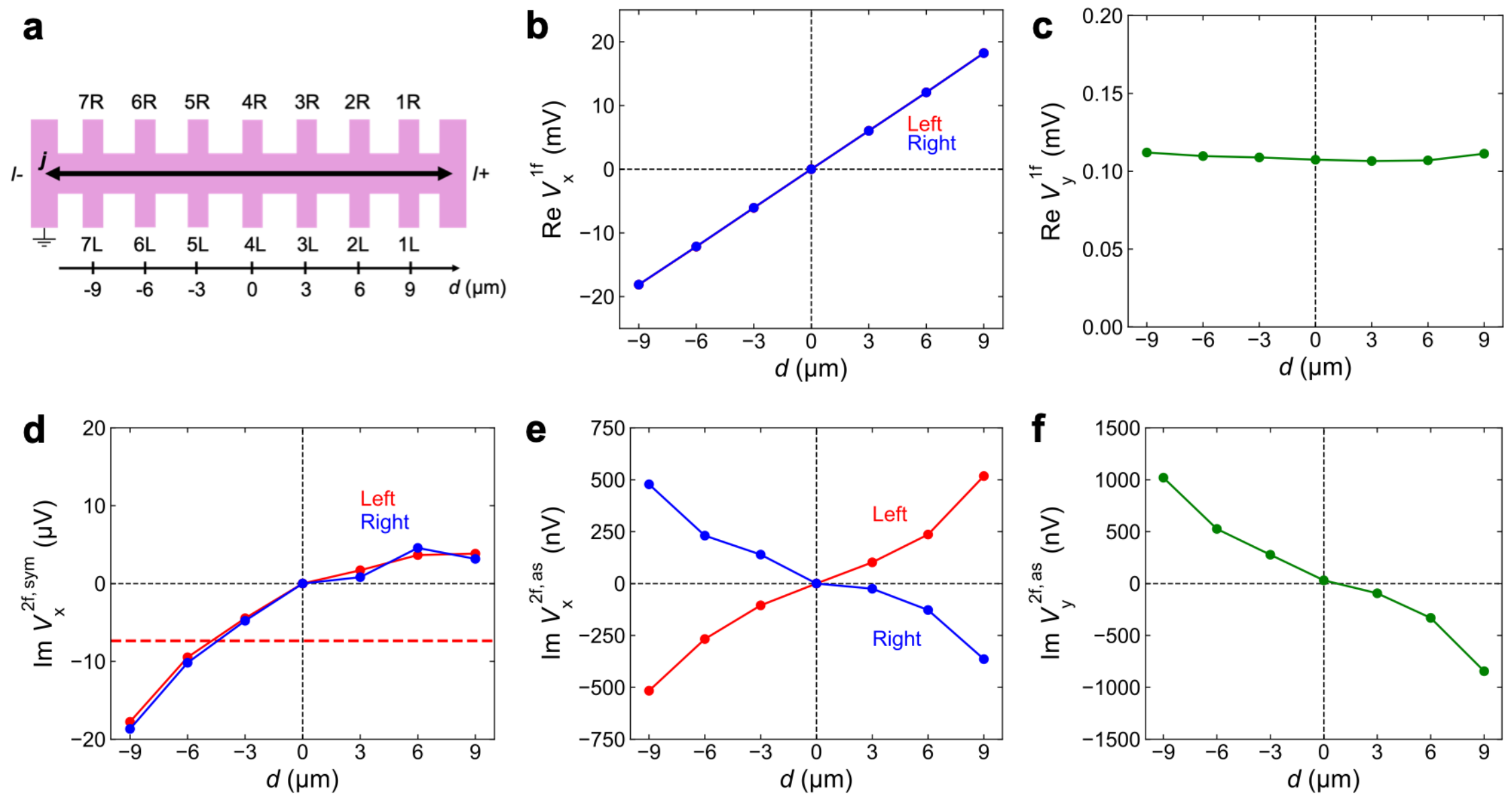

**Supplementary Fig. S9: Extended second-harmonic voltage mapping in a microdevice of $TbMn_6Sn_{4.4}In_{1.6}$ at 260 K.**

**a**, Schematic device structure of #4 in Fig. S2c. The ac current density, $j = 8.3 \times 10^8$ Am$^{-2}$ at $f = 513$ Hz, is applied along the device. The increased number of voltage terminals are labelled by numbers 1-7 and L (left side) and R (right side). The 1$f$ and 2$f$ longitudinal voltage is measured on both sides of the device at $n$L-($n$+1)L and $n$R-($n$+1)R ($n$ = 1, 2, …, 6) simultaneously. The 1$f$ and 2$f$ transverse voltage is also measured at $n$R-$n$L ($n$ = 1, 2, …, 7). **b**, 1$f$ longitudinal potential mapping on each side of the device. **c**, 1$f$ transverse potential mapping. **d**, **e**, 2$f$ longitudinal potential mapping of field-symmetric (**d**) and antisymmetric (**e**) components. Values in (**b**), (**d**), (**e**) are the potential measured from 4L or 4R. Red horizontal dashed line in (**d**) shows the central value between 1L(R) and 7L(R), which is shifted away from the positional centre (terminal 4). **f**, field-antisymmetric part of 2$f$ transverse potential ($n$R-$n$L), which is consistent with the 2$f$ longitudinal potential distribution in (**e**).

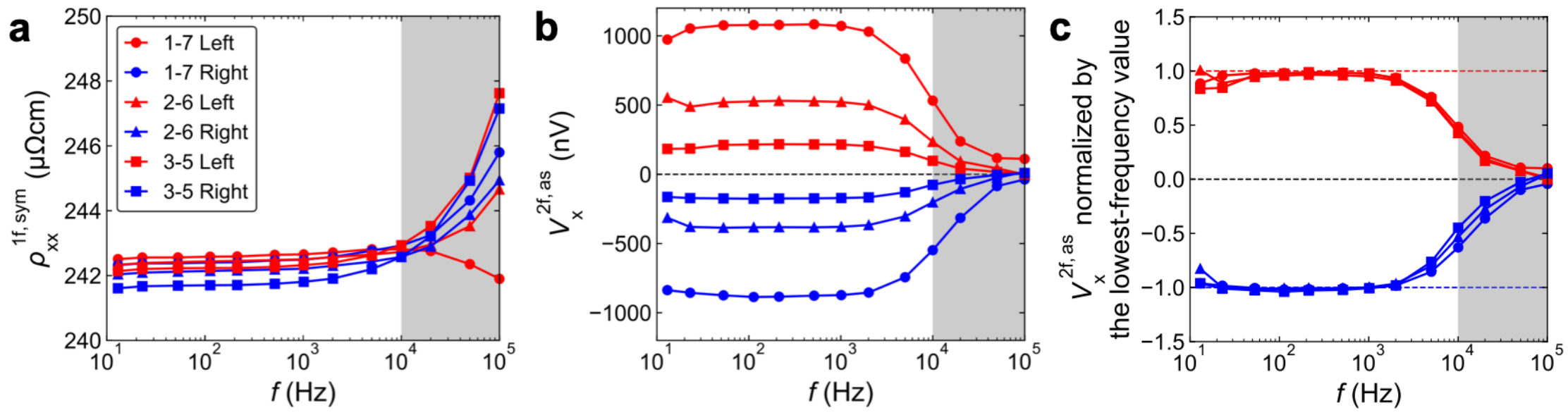

**Supplementary Fig. S10: Frequency dependence of boundary nonreciprocity at 260 K.**

**a, b**, Frequency dependence of longitudinal resistivity (a) and second-harmonic voltage (b) measured on both sides (red: left, blue: right) of the device #4 in Fig. S2c. The circle, triangle and square symbols indicate the voltage terminals used: 1-7, 2-6, and 3-5 (Fig. S3a), respectively. **c**, Data in (**b**) is normalized with the values at lowest frequency. The ac current density $j = 8.3 \times 10^8$ Am$^{-2}$ is applied. Both signals are independent of frequency up to 10 kHz. Above 10 kHz, the longitudinal resistivity starts to slightly change its magnitude, which indicates there is a limit where the ac electric current can be properly induced into the microdevice. The frequency dependence of the second harmonic signal is identical, regardless of the terminal pairs used.

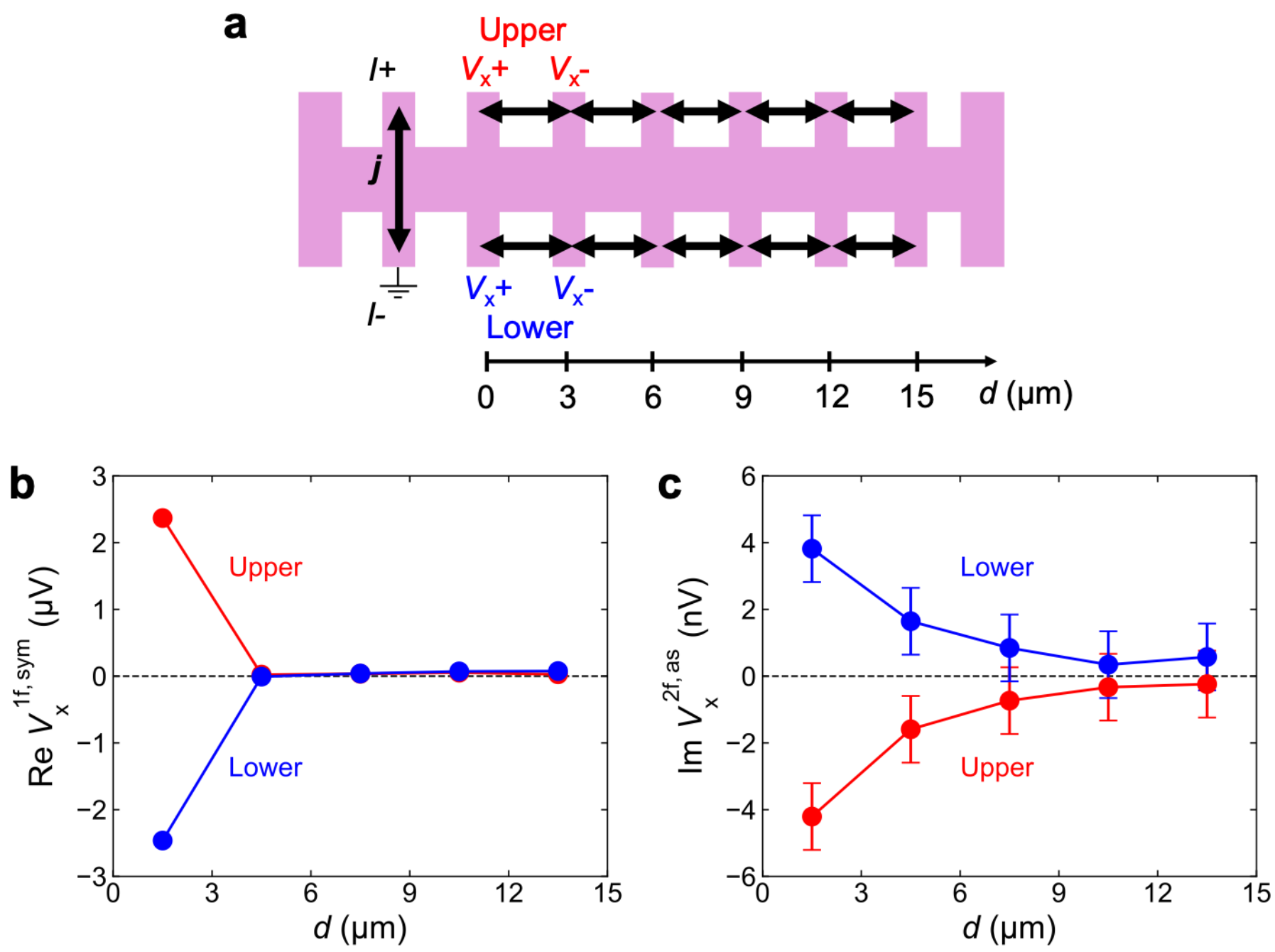

**Supplementary Fig. S11: Dissipation of boundary nonreciprocity.**

**a**, Schematic measurement configuration. The ac electric current density, $j = 8.3 \times 10^8$ $\mathrm{Am}^{-2}$ at $f = 513$ Hz, is applied using the voltage terminal 7 in Fig. S9a. The $1f$ and $2f$ longitudinal voltages are measured at each pair of remaining voltage terminals (black horizontal arrows) on both sides (upper and lower, simultaneously). **b**, **c**, Measured $1f$ (**b**) and $2f$ (**c**) voltages are plotted along the relative terminal position.

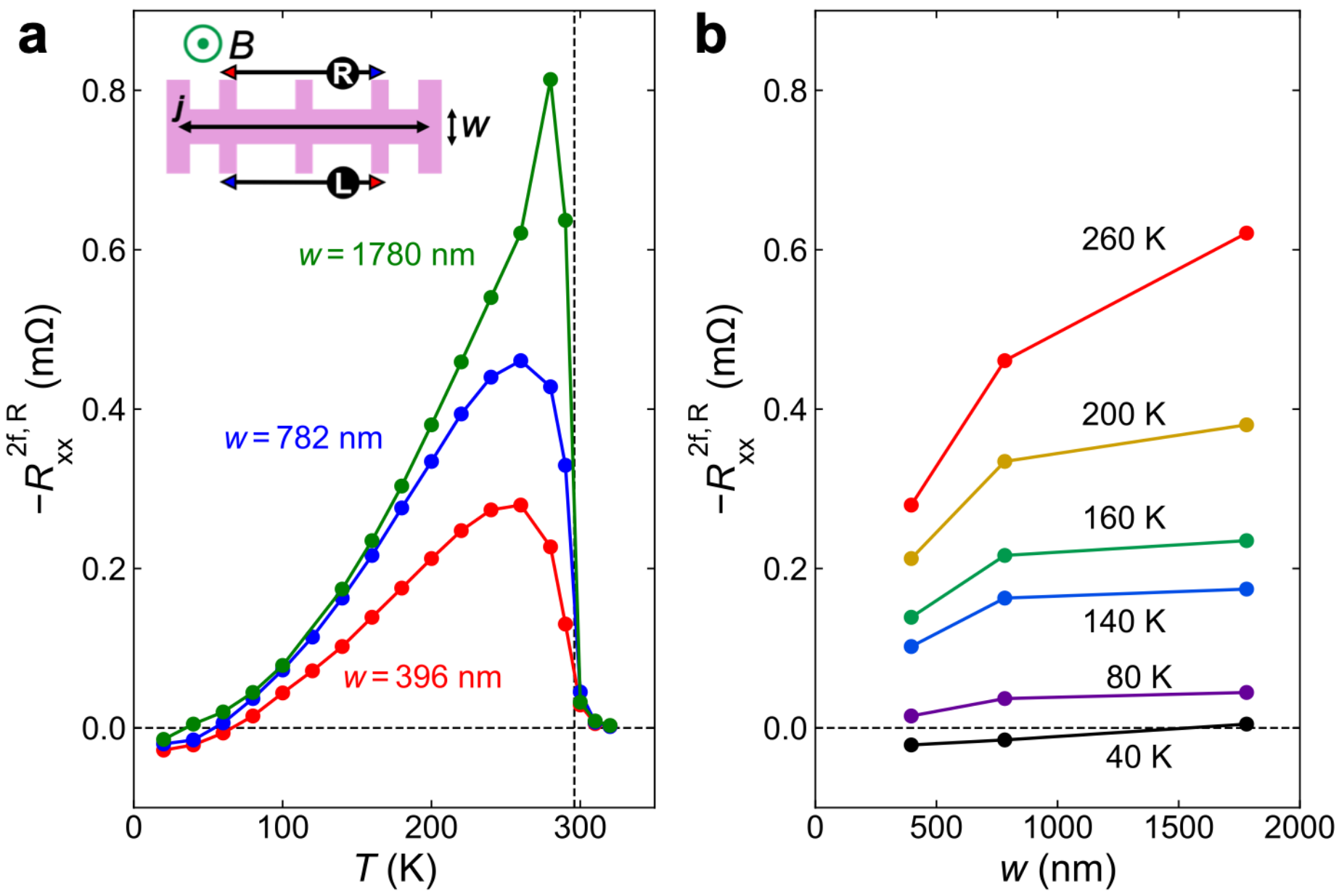

**Supplementary Fig. S12: Device-width dependence of the nonreciprocal resistance.** **a**, Temperature dependence of the nonreciprocal resistance in $B$ = 0 T measured on the right side of three FIB devices with different widths. The same ac current density, $j = 8.3 \times 10^8$ Am$^{-2}$ at $f$ = 513 Hz, is applied to each device. The inset defines the device width; all other dimensions are nominally identical (see Fig. S2a). The vertical dashed line in **a** indicates $T_C$. **b**, Device-width dependence of the nonreciprocal resistance in (**a**) at selected temperatures.